\documentclass{aa}
\usepackage{rotating}
\usepackage{graphicx}
\usepackage{times}
\begin{document}
\title{An {\it XMM-Newton} observation of the multiple system HD\,167971 (O5-8V + O5-8V + (O8I)) and the young open cluster NGC\,6604\thanks{Based on observations with XMM-Newton, an ESA Science Mission with instruments and contributions directly funded by ESA Member states and the USA (NASA). Partly based on observations collected at the European Southern Observatory (La Silla, Chile).}}
\author{M.\,De Becker\inst{1} \and G.\,Rauw\inst{1}\thanks{Research Associate FNRS (Belgium)} \and R.\,Blomme\inst{2} \and J.M.\,Pittard\inst{3} \and I.R.\,Stevens\inst{4} \and M.C.\,Runacres\inst{2}} 

\institute{Institut d'Astrophysique, Universit\'e de Li\`ege, All\'ee du 6 
Ao\^ut, B\^at B5c, B-4000 Li\`ege (Sart Tilman), Belgium \and
Royal Observatory of Belgium, Avenue Circulaire 3, 1180 Brussels, Belgium \and
Department of Physics \& Astronomy, University of Leeds, Leeds LS2 9JT, UK \and
School of Physics \& Astronomy, University of Birmingham, Edgbaston 
Birmingham B15 2TT, UK}

\date{Received date / Accepted date}
\authorrunning{M.\,De Becker et al.}
\titlerunning{An {\it XMM-Newton} observation of NGC\,6604}

\abstract{We discuss the results of two {\it XMM-Newton} observations of the open cluster NGC\,6604 obtained in April and September 2002. We concentrate mainly on the multiple system HD\,167971 (O5-8V + O5-8V + (O8I)). The soft part of the EPIC spectrum of this system is thermal with typical temperatures of about 2\,$\times$\,10$^{6}$ to 9\,$\times$\,10$^{6}$\,K. The nature (thermal vs non-thermal) of the hard part of the spectrum is not unambiguously revealed by our data. If the emission is thermal, the high temperature of the plasma ($\sim$ 2.3\,$\times$\,10$^{7}$ to 4.6\,$\times$\,10$^{7}$ K) would be typical of what should be expected from a wind-wind interaction zone within a long period binary system. This emission could arise from an interaction between the combined winds of the O5-8V + O5-8V close binary system and that of the more distant O8I companion. Assuming instead that the hard part of the spectrum is non-thermal, the photon index would be rather steep ($\sim$\,3). Moreover, a marginal variability between our two {\it XMM-Newton} pointings could be attributed to an eclipse of the O5-8V + O5-8V system. The overall X-ray luminosity points to a significant X-ray luminosity excess of about a factor 4 possibly due to colliding winds. Considering HD\,167971 along with several recent X-ray and radio observations, we propose that the simultaneous observation of non-thermal radiation in the X-ray (below 10.0\,keV) and radio domains appears rather unlikely. Our investigation of our {\it XMM-Newton} data of NGC\,6604 reveals a rather sparse distribution of X-ray emitters. Including the two bright non-thermal radio emitters HD\,168112 and HD\,167971, we present a list of 31 X-ray sources along with the results of the cross-correlation with optical and infrared catalogues. A more complete spectral analysis is presented for the brightest X-ray sources. Some of the members of NGC\,6604 present some characteristics suggesting they may be pre-main sequence star candidates.
}
\maketitle

\section{Introduction}

The nature of the X-ray emission from hot stars has been the subject of many discussions over the last three decades. The bulk of the X-ray emission of single O-stars arises in a hot optically thin thermal plasma inside the stellar wind. These plasmas are believed to be heated up to several 10$^6$\,K by shocks resulting from instabilites related to the line driving mechanism responsible for the mass loss of massive stars (see e.g. Owocki \& Rybicki \cite{OR}; Feldmeier et al.\,\cite{Feld}). The case of massive binaries is more complex, because they harbour a wind interaction zone that can produce a substantial additional X-ray emission (Stevens et al.\,\cite{SBP}). Such a wind collision is expected to heat the shocked plasma to temperatures of several 10$^7$\,K.

Some massive stars are known to display non-thermal emission in the radio domain (see e.g. Bieging et al.\,\cite{BAC}; Williams \cite{Wil}). Dougherty \& Williams (\cite{DW}) showed that most of the non-thermal radio emitting Wolf-Rayet (WR) stars are binaries, suggesting that the non-thermal phenomenon might be intimately related to multiplicity. However, the situation for O-stars is much less clear as the binary fraction among O-type non-thermal radio emitters is apparently lower than for WR stars (see Rauw \cite{rauwhk}).

The non-thermal radio emission, which is thought to be synchrotron emission (White \cite{Wh}), requires the presence of a population of relativistic electrons inside the radio emitting region, as well as the existence of a moderate magnetic field. On the one hand, high energy electrons could be accelerated through the first order Fermi mechanism in shocks (Pollock \cite{Pol}; Chen \& White \cite{CW}; Eichler \& Usov \cite{EU}). On the other hand, the existence of magnetic fields of a few hundred Gauss has recently been confirmed in the case of a few massive stars (Donati et al.\,\cite{Don1}, \cite{Don2}). For a discussion of the physical processes involved in this scenario, see e.g. De Becker et al. (\cite{HK}). If the ingredients required for non-thermal radio emission are present, one may wonder whether the relativistic electrons could produce a signature at other, essentially higher, energies. Indeed, the intense UV flux from massive stars could interact with relativistic electrons through Inverse Compton (IC) scattering resulting in a power law X-ray emission component (Eichler \& Usov \cite{EU}; Chen \& White \cite{CW2}). Such a scenario could apply both to single and binary systems, since both types of objects are believed to harbour hydrodynamic shocks that could accelerate particles. There are two major questions to be addressed in order to fully understand non-thermal phenomena in early-type stars: (1) one has to establish whether non-thermal X-ray emission really occurs, and (2) one has to check whether single stars are indeed able to produce non-thermal radiation.\\

In the framework of these considerations, non-thermal radio emitting massive stars are a priori privileged targets to search for an X-ray counterpart to this non-thermal radio emission. We therefore initiated a joint X-ray and radio campaign aiming at a better understanding of the emission processes in non-thermal radio emitters. The first target of this campaign observed with {\it XMM-Newton} was the O4V((f$^+$)) star 9\,Sgr (Rauw et al.\,\cite{9sgr}). Whilst the X-ray spectrum of 9\,Sgr displays indeed a hard emission tail, the very nature of the latter, thermal or non-thermal, could not be fully established. Optical spectra of 9\,Sgr revealed long term radial velocity variations suggesting that 9\,Sgr could indeed be a long-period binary system. The second object considered was HD\,168112 (De Becker et al.\,\cite{paper1}). This O5.5III(f$^+$) star, belonging to the NGC\,6604 cluster (Barbon et al.\,\cite{Bar}), also shows some evidence for binarity although no radial velocity variations were found. Indeed, De Becker et al. (\cite{paper1}) report changes of the nature of the radio emission (going from composite to strongly non-thermal, with a flux variation by a factor 5-7) simultaneously with a significant decrease of the X-ray flux (about 30\%), suggesting that a wind collision occurs in an eccentric long period binary system seen under low inclination. Blomme et al.\,(\cite{radio}) suggest a $\sim$\,1.4\,yr period for the radio fluxes. These results lend further support to the idea that binarity might be a necessary condition for non-thermal radio emission, also in O-stars. This paper is mainly devoted to the study of HD\,167971, another non-thermal radio emitter which is known to be a multiple system (Leitherer et al.\,\cite{triple}), and belongs also to NGC\,6604.\\ 

HD\,167971 was first classified as an O8f star (Hiltner \cite{Hil}). This system has been a target of several photometric studies (i.e. Johnson \cite{Joh}; Moffat \& Vogt \cite{MF}; Leitherer \& Wolf \cite{LW}). The IR study of Bertout et al. (\cite{Ber}) yielded ${\dot{\rm M}} \sim 9.4\,\times\,10^{-6}$\,M$_\odot$\,yr$^{-1}$ and v$_\infty \sim$ 2120\,km\,s$^{-1}$. In a more recent photometric and spectrometric study, Leitherer et al.\,(\cite{triple}) proposed this system to consist of a close eclipsing binary made up of 2 similar O-stars, with a third more luminous and more distant O-type companion (O5-8V + O5-8V + (O8I)). Their analysis of the light curve revealed a 3.3213 days period for the eclipsing binary. They found a mass loss rate of $\sim$\,2\,$\times$\,10$^{-6}$\,M$_\odot$\,yr$^{-1}$ and a terminal velocity of 3100\,km\,s$^{-1}$ for the brightest third star. The study of Davidge \& Forbes (\cite{DF}) confirmed the orbital period of the close binary system. Their photometric solution, which is in better agreement with observations, has non-zero third light, suggesting that HD\,167971 is indeed a multiple system. However, we note that at this stage there is no evidence of a gravitational link between the third star and the eclipsing binary.

In his review of radio emitting hot stars, Williams (\cite{Wil}) lists HD\,167971 as a non-thermal emitter, with a radio spectral index of $-$0.6 and a 6\,cm flux level ranging between 13.8 and 17.0\,mJy. It is the strongest non-thermal radio emitter reported in the study of Bieging et al. (\cite{BAC}). This star appears consequently to be a good candidate to study the implication of binarity on non-thermal emission processes. This paper is devoted to the analysis of two {\it XMM-Newton} pointings obtained in 2002. The radio results from the VLA, as well as a discussion of archive radio data, will be presented in a separate paper (Blomme et al.\,\cite{Blom2}, in preparation).\\

NGC\,6604 is a rather compact open cluster lying at the core of the HII region S54 (Georgelin et al.\,\cite{geor}), in the Ser\,OB2 association (Forbes\,\cite{for}). The study of Forbes \& DuPuy (\cite{FD}) revealed that it is rather young ($\sim$\,4\,Myr) as suggested also by the possible presence of pre-main sequence (PMS) objects. However, the study of Barbon et al.\,(\cite{Bar}) did not confirm the existence of pre-main sequence stars, although these authors also derived an age of the cluster of about 5\,$\pm$\,2\,Myr. The same authors derive a distance to the cluster of about 1.7\,kpc. In addition to the study of the two non-thermal radio emitters mentioned hereabove, the X-ray observation of NGC\,6604 offers also the possibility to study the stellar population of the cluster. Indeed, recent {\it XMM-Newton} observations of NGC\,6383 (Rauw et al.\,\cite{ngc6383}) and NGC\,6231 (Sana et al.\,\cite{ngc6231}) reveal a strong concentration of X-ray selected PMS stars around the most massive cluster members that are found in the cluster core. Such mass segregated clusters with a concentration of low-mass PMS objects could indicate that the most massive stars have formed through a combination of gas accretion and stellar collisions in the very dense core of the cluster (Bonnell et al.\,\cite{BBZ}; Bonnell \& Bate\,\cite{BB}; Bonnell et al.\,\cite{BBV}). The {\it XMM-Newton} observation of NGC\,6604 is expected to bring new insight on the existence of low-mass PMS stars in this cluster.\\

This paper is organized as follows. In Sect.\,\ref{sect_obs}, we briefly discuss the data reduction. Section\,\ref{sect_epic2} describes the X-ray spectral analysis, as well as some considerations relevant to the colliding winds context. In Sect.\,\ref{arch} we compare our results for HD\,167971 with archival X-ray observations. Section\,\ref{sect_disc} is devoted to a general discussion of HD\,167971. The X-ray emission from the other sources in the field of view is discussed in Sect.\,\ref{ngc}. Finally, our conclusions are presented in Sect.\,\ref{sect_concl}.

\section{Observations \label{sect_obs}}
We obtained two {\it XMM-Newton} (Jansen et al.\,\cite{xmm}) observations of the NGC\,6604 open cluster during revolution 426, in April 2002 (Obs.ID\,0008820301, JD\,2\,452\,372.477 -- 2\,452\,372.637), and revolution 504, in September 2002 (Obs.ID\,0008820601, JD\,2\,452\,526.694 -- 2\,452\,526.868). The aim-point of these observations was set in such a way as to monitor simultaneously the two non-thermal radio emitters HD\,168112 and HD\,167971 within a single EPIC field of view. The exposure time for both observations was about 13\,ks. For details on these exposures and on the reduction procedure, we refer to the paper devoted to HD\,168112 (De Becker et al.\,\cite{paper1}).

\begin{figure*}[ht]
\begin{center}
\resizebox{17.5cm}{9.0cm}{\includegraphics{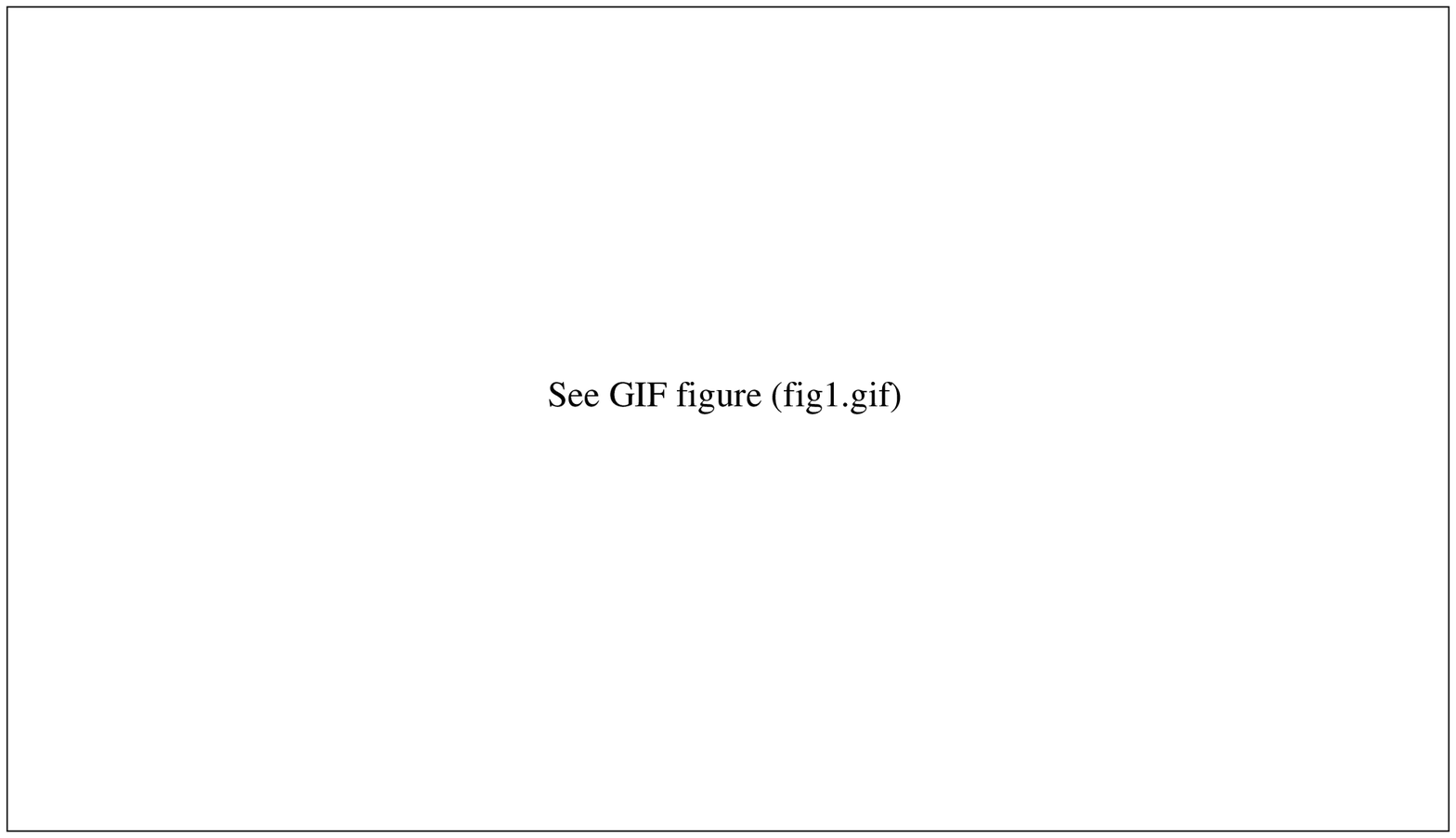}}
\caption{Source (circle) and background (annulus) regions selected for the spectrum extraction of HD\,167971. Boxes were used to exclude the CCD gaps and bad columns. We see clearly that some observations are strongly affected by the presence of these gaps. The images are displayed in detector coordinates.\label{reg2}}
\end{center}
\end{figure*}

The X-ray events of HD\,167971 were selected from inside a 60 arcsec radius circular region centered on the star. The background was extracted from an annular region around the source. The inner and outer radius of this annulus were chosen to obtain a surface area roughly equal to the source region area. Both regions are shown in Fig.\,\ref{reg2} in the case of the three EPIC fields, for both observations. In each case, the source region is affected by at least one gap, which was excluded by means of a rectangular box. These boxes were adjusted following a careful inspection of the adequate exposure maps. Unfortunately these gaps sometimes cross the source region close to its center, strongly affecting the results discussed hereafter. In the case of the EPIC-pn data sets, the background region is crossed by a bad column which was also removed with a box.

\section{The EPIC spectrum of HD\,167971 \label{sect_epic2}}
\subsection{Spectral analysis\label{specan}}

As already stated by De Becker et al. (\cite{paper1}), the end of the April 2002 exposure is affected by a strong soft proton flare (see Lumb \cite{lumb}). However, it appears that the background subtraction provides an efficient correction, and the contaminated time interval was therefore not rejected. Indeed, the level of the background is at least a factor 2 lower than the mean count rate inside the source region\footnote{In the case of HD\,168112 (De Becker et al. \cite{paper1}), the soft proton flare did not affect significantly the spectral analysis although the background level was of the same order of magnitude as the source count rate.}.
 
To carry out this spectral analysis, we followed the same procedure as already applied to other targets like HD\,159176 (De Becker et al.\,\cite{DeB}) and HD\,168112 (De Becker et al.\,\cite{paper1}). We chose a list of models including thermal and non-thermal emission components. Massive stars are known to display thermal spectra with strong and broad emission lines (e.g. $\zeta$\,Pup, Kahn et al.\,\cite{Kahn}; $\tau$\,Sco, Cohen et al.\,\cite{Coh}). Most current models consider that, in the case of single O-stars, such hot plasmas are produced by strong shocks inside the stellar wind (Feldmeier et al.\,\cite{Feld}). To model the emission spectrum arising from such plasmas, we used an optically thin thermal plasma {\tt mekal} model (Mewe et al.\,\cite{mewe}; Kaastra \cite{ka}) available within {\sc xspec}. In the case of massive binaries, colliding winds can also heat the plasma in the wind interaction zone to a temperature of up to a few times 10$^{7}$ K. The emission from the wind interaction zone can also be modelled with optically thin thermal plasma codes (see e.g. the case of HD\,159176, De Becker et al.\,\cite{DeB}). Moreover, to account for the possibility of non-thermal emission arising from the Inverse Compton (IC) scattering of UV photons due to the presence of a population of relativistic electrons (Pollock \cite{Pol}; Chen \& White \cite{CW}; Eichler \& Usov \cite{EU}), power law models were also considered. Our grid of models therefore includes pure thermal models, and thermal plus non-thermal models.

The $\chi^2$ minimization technique was used to estimate the quality of the fits described in the following paragraphs of this section. The parameter values quoted in Tables \ref{fitTTT} and \ref{fitTTP} are obtained through this method. However, to check the consistency of our approach, we confronted these results with those obtained with other techniques. Indeed, when one deals with small count numbers the parameter estimation could be achieved through the C-statistic (Cash \cite{cash}), or alternatively with the $\chi^2$ statistic using a Churazov weighting (Churazov et al. \cite{chur}). Both methods yield results compatible with those quoted in Tables \ref{fitTTT} and \ref{fitTTP} within the error bars.\\

\subsubsection{Wind absorption \label{abs}}

In each case, two separate absorption columns were used to account for interstellar (ISM) and circumstellar (wind material) absorption respectively. The first absorption component is due to neutral material and the hydrogen column density was fixed at 0.63\,$\times$\,10$^{22}$ cm$^{-2}$. This value was obtained through the gas to dust ratio given by Bohlin et al. (\cite{Boh}), using a $(B - V)$ color index of +0.77 (Chlebowski et al.\,\cite{Chle}) and an intrinsic color index $(B - V)_o$ of --0.31 estimated for an O8 star (see e.g. Mihalas \& Binney \cite{MB}). For the second absorption column, left as a free parameter while fitting the spectra, we used the ionized wind model already presented by Naz\'e et al.\,(\cite{108}). This model considers the absorption by the 10 most abundant elements (H, He, C, N, O, Ne, Mg, Si, S and Fe) fixed at their solar abundances (Anders \& Grevesse\,\cite{AG}). We used the stellar fluxes from the Kurucz library of spectra, and a standard wind velocity law with $\beta$\,=\,0.8.

\begin{figure}[ht]
\begin{center}
\resizebox{8.5cm}{5.0cm}{\includegraphics{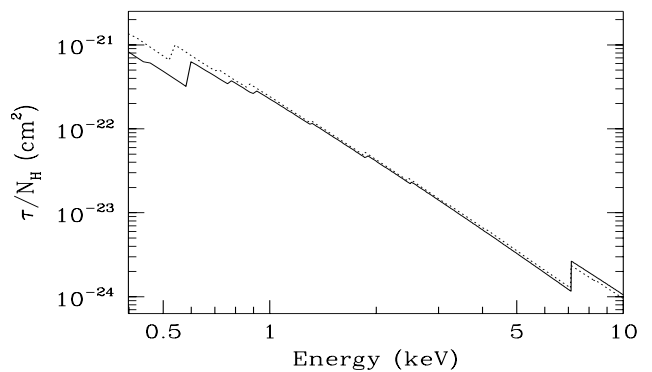}}
\caption{Opacity resulting from the normalization of the optical depth by the hydrogen column density as a function of the energy for ionized (solid line) and neutral (dashed line) wind material.\label{opa}}
\end{center}
\end{figure}

The main problem to model the wind properties of HD\,167971 with this model is that it only accounts for the geometry of a single star wind, although we know that our target is a triple system. The choice of the parameters adequate for our purpose is therefore not straightforward. To check the sensitivity of the wind opacity calculated by this model regarding the stellar parameters, we computed several opacity tables covering a parameter space typical for a range of spectral types including that of the components of HD\,167971. This grid was calculated for mass loss rates ranging from 10$^{-6}$ to 10$^{-5}$\,M$_\odot$\,yr$^{-1}$, and terminal velocities between 2200 and 3100\,km\,s$^{-1}$. These domains cover different values proposed by some authors for the parameters of HD\,167971 (Bertout et al.\,\cite{Ber}; Leitherer et al.\,\cite{triple}). We used also three different Kurucz spectra, respectively referred to as {\it low} ($T_\mathrm{eff}$=30\,000; log\,$g$=3.5), {\it medium} ($T_\mathrm{eff}$=35\,000; log\,$g$=4.0), and {\it high} ($T_\mathrm{eff}$=40\,000; log\,$g$=4.5), and we computed the optical depth ($\tau$) as a function of the energy for X-ray emitting shells located between 1.5 and 100 stellar radii. The opacities resulting from the normalization of $\tau$ by the hydrogen column of the wind is very similar whatever the case considered within this parameter space. Consequently, any set of parameters located within the boundaries discussed hereabove appears to be adequate. In the next section, we will discuss spectral fittings using the opacities we obtained from the {\it high} Kurucz spectrum, and the parameters quoted in Table\,\ref{param}. The parameters were selected on the basis of the values used or derived by Leitherer et al.\,(\cite{triple}) in the computation of synthetic H$\alpha$ profiles for this star. Fig.\,\ref{opa} shows the normalized $\tau$ as a function of the energy for the ionized wind (solid line), along with the same quantity for neutral material (dashed line). We see that the largest deviation between the neutral and ionized cases is observed at low energies. The opacity table obtained with the parameters listed in Table\,\ref{param} was converted into a FITS table in a format suitable to be used as a multiplicative absorption model within the {\sc xspec} software. It will be referred to as the {\tt wind} absorption model in the remainder of this paper.

\begin{table}
\caption{Parameters selected to compute the opacity table used for the modelling of the absorption of X-rays by the ionized wind material.\label{param}}
\begin{center}
\begin{tabular}{lc}
\hline
\hline
\vspace*{-0.3cm}\\
$T_\mathrm{eff}$ (K) & 35\,000 \\
$\log{g}$ & 4.0 \\
${\dot{\rm M}}$ (M$_\odot$\,yr$^{-1}$) & 5\,$\times$\,10$^{-6}$ \\
$V_\infty$ (km\,s$^{-1}$) & 3100 \\
$R_{shell}$ (R$_*$) & 5 \\
\vspace*{-0.3cm}\\
\hline
\end{tabular}
\end{center}
\end{table}

\subsubsection{Spectral fittings\label{res}}

\begin{table*}[ht]
\caption{Parameters for EPIC spectra of HD\,167971 in the case of a {\tt wabs*wind*(mekal+mekal+mekal)} model. Results are given for MOS1, MOS2, combined MOS in the case of the April observation, and for EPIC-pn only in the case of the September one. The first absorption component is frozen at 0.63\,$\times$\,10$^{22}$ cm$^{-2}$. The second absorption column, quoted as $N_\mathrm{w}$ (in cm$^{-2}$), stands for the absorption by the ionized wind material. The last column gives the observed flux between 0.4 and 10.0 keV. The normalization parameter (Norm) of the {\tt mekal} component is defined as $(10^{-14}/(4\,\pi\,D^2))\int{n_\mathrm{e}\,n_\mathrm{H}\,dV}$, where $D$, 
$n_\mathrm{e}$ and $n_\mathrm{H}$ are respectively the distance to the source (in cm), and the electron and hydrogen number densities (in cm$^{-3}$). The error bars represent the 1-$\sigma$ confidence interval.\label{fitTTT}}
\begin{center}
\begin{tabular}{cccccccccc}
\hline
\hline
\vspace*{-0.3cm}\\
	& Log\,$N_\mathrm{w}$ & $kT_1$ & Norm$_1$ & $kT_2$ & Norm$_2$ & $kT_3$ & Norm$_3$ & $\chi^2_\nu$ & Obs.Flux \\
	&  & (keV) &  & (keV) &  & (keV) &  & d.o.f. & (erg\,cm$^{-2}$\,s$^{-1}$) \\
\hline
\vspace*{-0.3cm}\\
\multicolumn{10}{l}{April}\\
\vspace*{-0.3cm}\\
\hline
\vspace*{-0.3cm}\\
MOS1	& 21.77 & 0.27 & 1.38\,$\times$\,10$^{-2}$ & 0.79 & 2.05\,$\times$\,10$^{-3}$ & 3.30 & 6.38\,$\times$\,10$^{-4}$ & 0.867 & 1.79\,$\times$\,10$^{-12}$\\ 
	& $\pm$ 0.08 & $\pm$ 0.02 & $\pm$ 0.67\,$\times$\,10$^{-2}$ & $\pm$ 0.08 & $\pm$ 0.72\,$\times$\,10$^{-3}$ & $\pm$ 1.32 & $\pm$ 3.86\,$\times$\,10$^{-4}$ & 120 & \\
\hline
\vspace*{-0.3cm}\\
MOS2	& 21.78 & 0.24 & 1.68\,$\times$\,10$^{-2}$ & 0.77 & 2.62\,$\times$\,10$^{-3}$ & 3.96 & 4.26\,$\times$\,10$^{-4}$ & 1.115 & 1.69\,$\times$\,10$^{-12}$\\ 
	& $\pm$ 0.07 & $\pm$ 0.02 & $\pm$ 0.76\,$\times$\,10$^{-2}$ & $\pm$ 0.06 & $\pm$ 0.76\,$\times$\,10$^{-3}$ & $\pm$ 2.68 & $\pm$ 3.81\,$\times$\,10$^{-4}$ & 116 & \\
\hline
\vspace*{-0.3cm}\\
MOS1	& 21.78 & 0.25 & 1.52\,$\times$\,10$^{-2}$ & 0.78 & 2.35\,$\times$\,10$^{-3}$ & 3.57 & 5.21\,$\times$\,10$^{-4}$ & 0.973 & 1.70\,$\times$\,10$^{-12}$\\ 
+ MOS2	& $\pm$ 0.05 & $\pm$ 0.01 & $\pm$ 0.50\,$\times$\,10$^{-2}$ & $\pm$ 0.05 & $\pm$ 0.52\,$\times$\,10$^{-3}$ & $\pm$ 1.25 & $\pm$ 2.73\,$\times$\,10$^{-4}$ & 243 & \\
\hline
\vspace*{-0.3cm}\\
\multicolumn{10}{l}{September}\\
\vspace*{-0.3cm}\\
\hline
\vspace*{-0.3cm}\\
pn	& 21.69 & 0.27 & 9.55\,$\times$\,10$^{-3}$ & 0.77 & 1.87\,$\times$\,10$^{-3}$ & 2.40 & 5.47\,$\times$\,10$^{-4}$ & 1.089 & 1.49\,$\times$\,10$^{-12}$\\ 
	& $\pm$ 0.06 & $\pm$ 0.01 & $\pm$ 2.86\,$\times$\,10$^{-3}$ & $\pm$ 0.05 & $\pm$ 0.43\,$\times$\,10$^{-3}$ & $\pm$ 0.43 & $\pm$ 2.32\,$\times$\,10$^{-4}$ & 244 & \\
\hline
\end{tabular}
\end{center}
\end{table*}

\begin{table*}[ht]
\caption{Same as Table\,\ref{fitTTT}, but in the case of a {\tt wabs*wind*(mekal+mekal+power)} model. For the power law component, the normalization parameter (Norm$_3$) corresponds to the photon flux at 1 keV.\label{fitTTP}}
\begin{center}
\begin{tabular}{cccccccccc}
\hline
\hline
\vspace*{-0.3cm}\\
	& Log\,$N_\mathrm{w}$ & $kT_1$ & Norm$_1$ & $kT_2$ & Norm$_2$ & $\Gamma$ & Norm$_3$ & $\chi^2_\nu$ & Obs.Flux\\
	&  & (keV) &  & (keV) &  &  &  & d.o.f. & (erg\,cm$^{-2}$\,s$^{-1}$) \\
\hline
\vspace*{-0.3cm}\\
\multicolumn{10}{l}{April}\\
\vspace*{-0.3cm}\\
\hline
\vspace*{-0.3cm}\\
MOS1	& 21.48 & 0.31 & 4.24\,$\times$\,10$^{-3}$ & 0.75 & 1.36\,$\times$\,10$^{-3}$ & 3.04 & 8.47\,$\times$\,10$^{-4}$ & 0.851 & 1.70\,$\times$\,10$^{-12}$\\ 
	& $\pm$ 0.28 & $\pm$ 0.05 & $\pm$ 5.77\,$\times$\,10$^{-3}$ & $\pm$ 0.08 & $\pm$ 0.78\,$\times$\,10$^{-3}$ & $\pm$ 0.282 & $\pm$ 5.09\,$\times$\,10$^{-4}$ & 120 & \\
\hline
\vspace*{-0.3cm}\\
MOS2	& 21.66 & 0.25 & 9.54\,$\times$\,10$^{-3}$ & 0.75 & 2.04\,$\times$\,10$^{-3}$ & 2.95 & 6.07\,$\times$\,10$^{-4}$ & 1.095 & 1.63\,$\times$\,10$^{-12}$\\ 
	& $\pm$ 0.14 & $\pm$ 0.02 & $\pm$ 6.66\,$\times$\,10$^{-3}$ & $\pm$ 0.06 & $\pm$ 0.94\,$\times$\,10$^{-3}$ & $\pm$ 0.426 & $\pm$ 6.25\,$\times$\,10$^{-4}$ & 116 & \\
\hline
\vspace*{-0.3cm}\\
MOS1	& 21.63 & 0.27 & 8.02\,$\times$\,10$^{-3}$ & 0.76 & 1.78\,$\times$\,10$^{-3}$ & 2.94 & 6.78\,$\times$\,10$^{-4}$ & 0.954 & 1.70\,$\times$\,10$^{-12}$\\ 
+ MOS2	& $\pm$ 0.11 & $\pm$ 0.02 & $\pm$ 4.65\,$\times$\,10$^{-3}$ & $\pm$ 0.05 & $\pm$ 0.64\,$\times$\,10$^{-3}$ & $\pm$ 0.268 & $\pm$ 4.21\,$\times$\,10$^{-4}$ & 243 & \\
\hline
\vspace*{-0.3cm}\\
\multicolumn{10}{l}{September}\\
\vspace*{-0.3cm}\\
\hline
\vspace*{-0.3cm}\\
pn	& 21.68 & 0.27 & 9.08\,$\times$\,10$^{-3}$ & 0.78 & 1.84\,$\times$\,10$^{-3}$ & 2.51 & 2.96\,$\times$\,10$^{-4}$ & 1.072 & 1.52\,$\times$\,10$^{-12}$\\ 
	& $\pm$ 0.07 & $\pm$ 0.01 & $\pm$ 3.18\,$\times$\,10$^{-3}$ & $\pm$ 0.05 & $\pm$ 0.47\,$\times$\,10$^{-3}$ & $\pm$ 0.33 & $\pm$ 2.78\,$\times$\,10$^{-4}$ & 244 & \\
\hline
\end{tabular}
\end{center}
\end{table*}

As previously mentioned, most of our data sets are strongly affected by the presence of gaps and/or bad columns. Only the EPIC-MOS1 and EPIC-MOS2 observations of April, and to some extent the EPIC-pn observation of September appear to be only slightly affected. In all other cases, the CCD gaps lie very close to the center of the PSF of the source. Since the data seriously affected by gaps yield unreliable spectral fits, we consider only the spectral fits for those data that are least affected.

Two-component models were used as a first step with some success, but it appeared that those models were unable to fit data above about 4-5 keV. As a consequence, we used three component models, each one having two {\tt mekal} thermal components, plus another thermal or power law component. Solar abundances are assumed throughout our fitting procedure. The results obtained with these two models are summarized in Tables \ref{fitTTT} and \ref{fitTTP}. The last column of these tables yields the observed (i.e. absorbed) flux integrated between 0.4 and 10.0\,keV for each set of model parameters. The dispersion on the flux was evaluated from the range of fluxes for the models with parameters covering the confidence intervals quoted in Tables \ref{fitTTT} and \ref{fitTTP}. According to this approach, the error on the flux in that energy band is expected to be about 5\,\%.\\

\begin{figure}[htb]
\begin{center}
\resizebox{8.5cm}{6.0cm}{\includegraphics{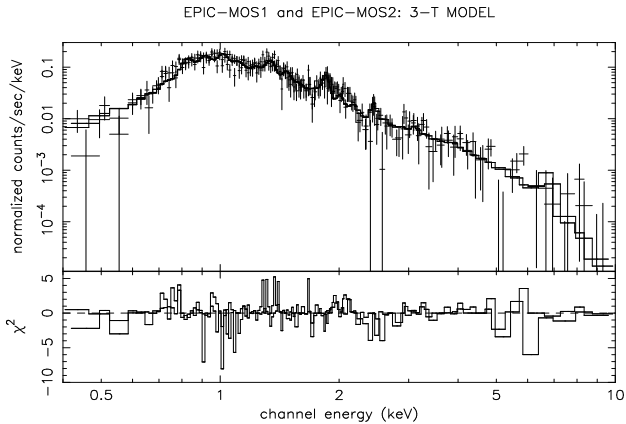}}
\caption{Combined EPIC-MOS1 and EPIC-MOS2 spectra of HD\,167971 for the April 2002 observation, fitted with a {\tt wabs*wind*(mekal+mekal+mekal)} model between 0.4 and 10.0 keV.\label{spmos}}
\end{center}
\end{figure}

\begin{figure}[htb]
\begin{center}
\resizebox{8.5cm}{6.0cm}{\includegraphics{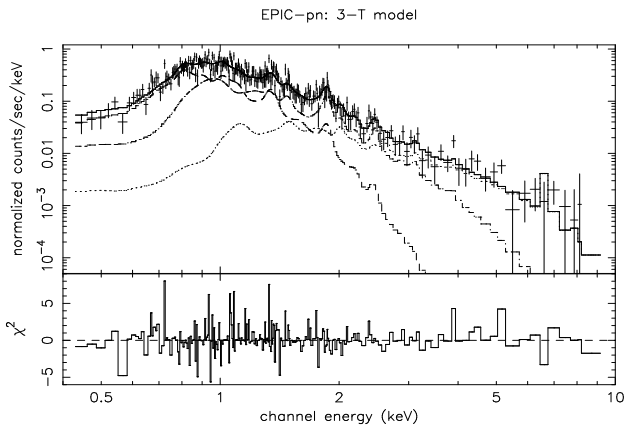}}
\caption{EPIC-pn spectrum of HD\,167971 of the September 2002 observation, fitted with a {\tt wabs*wind*(mekal+mekal+mekal)} model between 0.4 and 10.0 keV. The three components are individualy displayed.\label{sppn}}
\end{center}
\end{figure}

First, the EPIC-MOS1 and EPIC-MOS2 results of the April observation appear to be very consistent, and our models were also fitted to the two data sets simultaneously. A comparison between the results of the two models reveals no significant differences in the quality of the fit. Both models are nearly equivalent within the error bars for the two first thermal {\tt mekal} components. The $\chi_\nu^2$ values obtained and used as goodness-of-fit criterion are very similar for all EPIC instruments, either considered independently or combined. Moreover, the fluxes estimated on the basis of these two models are similar within the error bars, reinforcing the idea that both models reproduce the spectral shape rather consistently. Unfortunately, the poor quality of the data above 5 keV does not allow us to unambiguously check for the presence of the Fe K line that could help constrain the nature of the hard spectral component. These results show that the ambiguity of the nature of the hard part of the spectrum (thermal versus non-thermal) can not be alleviated with our data. This case is reminiscent of the results of the {\it XMM-Newton} observations obtained in the case of other non-thermal radio emitters (see e.g. 9\,Sgr, Rauw et al.\,\cite{9sgr}).

The EPIC-pn observation of September gives results very similar to those of the EPIC-MOS ones of April. Even if some systematic deviations could be expected as the EPIC-MOS and EPIC-pn instruments are somewhat different, we see that the overall spectral shape remained more or less steady between the two observations. If we consider the three-temperature model, we see that the emission measure of the soft thermal component decreases from April to September. This decrease is responsible for the lower flux quoted in Table\,\ref{fitTTT} for the September pointing. However, if we consider the model described in Table\,\ref{fitTTP}, it is the normalization factor of the power law which decreases between the two pointings whilst the parameters of the two thermal components remain steady.

\subsection{X-ray luminosity of HD\,167971 \label{lumin}}

\subsubsection{X-ray variability \label{var}}

When investigating the X-ray variability of a system like HD\,167971, one has to consider several scenarios likely to produce variations in the X-ray luminosity. Variability could indeed be expected for several reasons. First, the interaction of the stellar winds of the two components of the O5-8V + O5-8V close binary system could vary with the phase. Since the close binary has a circular orbit (van Genderen et al.\,\cite{vanG}), the only way to produce a modulation of the X-ray flux from the wind-interaction zone would come from a change in the column density along the line of sight as a function of the orbital phase. The resulting change in absorption would then mostly affect the soft part of the EPIC spectrum. Nevertheless, no significant change appears in the spectrum or in the absorption component of the model used to fit the data between the first and the second exposure. A second possible origin could be the putative wind-wind interaction due to the presence of the third star, if the system in indeed triple. The stellar wind of the third component is indeed likely to produce a second interaction zone by colliding with the winds of the stars of the close binary system. Finally, variability could possibly result from the eclipse of the O5-8V + O5-8V binary system. According to the ephemeris derived by Leitherer et al. (\cite{triple}), the April observation was performed between phases 0.64 and 0.69, and the September one between phases 0.04 and 0.12. As phase zero corresponds to the primary eclipse, we see that the September {\it XMM-Newton} observation was performed when the primary was partially hidden by the secondary, allowing us to possibly expect a lower X-ray flux than during the April observation when no eclipse occurred. However, the extrapolation of this ephemeris to the epoch of our observations must be considered with caution, especially for such a short period. To clarify this question, we investigated the X-ray variability of HD\,167971 on the basis of the results of our two {\it XMM-Newton} observations, along with a confrontation to archive data (see Sect.\,\ref{arch}).\\

Count rates of HD\,167971 between 0.4 and 10.0 keV are presented in Table\,\ref{lum2}. Since CCD gaps (see Fig.\,\ref{reg2}) affect the area of the effective source region considered for the event selection, the variability between our two observations can not be discussed on the basis of count rates.

\begin{table}
\caption{Observed count rates for the three EPIC instruments evaluated between 0.4 and 10.0 keV. The error bars on the count rates represent the 1-$\sigma$ confidence interval.\label{lum2}}
\begin{center}
\begin{tabular}{cccc}
\hline
\hline
Observation & MOS1 CR & MOS2 CR & pn CR \\
	& (cts\,s$^{-1}$) & (cts\,s$^{-1}$) & (cts\,s$^{-1}$) \\
\hline
\vspace*{-0.3cm}\\
April 2002 & 0.146 & 0.143 & -- \\
  & $\pm$ 0.004 & $\pm$ 0.004 &  \\
September 2002 & -- & -- & 0.465 \\
  &  &  & $\pm$ 0.008 \\
\vspace*{-0.3cm}\\
\hline
\end{tabular}
\end{center}
\end{table}

To obtain a more consistent, although model dependent way to quantify the X-ray emission, absorbed fluxes and fluxes corrected for interstellar absorption were evaluated on the basis of the pure thermal model. The fluxes quoted in Table\,\ref{fitTTT} give a rough idea of the behaviour of HD\,167971 between April and September 2002. If we make the assumption that we can compare the fluxes obtained separately with the EPIC-MOS and EPIC-pn instruments, we observe a slight decrease of the X-ray flux ($\sim$\,12\,\%). However, we remind that the error on the flux estimate is about 5\,\%, and that such a heterogeneous comparison may introduce another systematic error of a few percent. Therefore, the variability of HD\,167971 between the two {\it XMM-Newton} observations is of rather low level. If this slight variation is real, it can be due to the eclipse occurring during the second observation as pointed out above. We note however that the decrease of the emission measure is observed mostly for the soft component in the case of the three-temperature model, whilst the case of 
V444\,Cyg discussed by Pittard (\cite{Pit}) shows that the eclipse would indeed be expected to reduce mainly the flux in the hard part of the X-ray spectrum. A detailed modelling of the effect of the absorption by the winds, of the occultation of the colliding zone, and of the dynamic of the stellar winds including radiative inhibitaion effects is needed to investigate the effect of the eclipse on the observed X-ray spectrum.\\

The variability on shorter time scales was also investigated. Light curves were generated in different energy bands between 0.4 and 10.0\,keV, with time bins ranging from 100 s to 1000\,s. These light curves were background corrected, and each time bin accounts for Good Time Intervals (GTIs)\footnote{Even if no flare contaminated time interval was rejected, standard GTIs are anyway always applied to {\it XMM-Newton} data, and should be taken into account in every timing analysis.}. Variability tests (chi-square, {\it pov}-test\footnote{See Sana et al.\,(\cite{sana}) for a detailed discussion of this variability test.}) were applied to all of them and no significant short term variability was found.

\subsubsection{Overall luminosity \label{ol}}

To obtain an estimate of the overall X-ray luminosity corrected for the ISM absorption, let us consider the model parameters obtained in the case of EPIC-MOS for the first observation, and EPIC-pn for the second one. This yields an $L_\mathrm{X}$ of about 4.3$\times$\,10$^{33}$ and 3.8$\times$\,10$^{33}$\,erg\,s$^{-1}$ respectively for the April and September observations, for a distance of 2\,kpc. According to the bolometric magnitudes given by van Genderen et al. (\cite{vanG}) for the three components of this presumably triple system, we can infer bolometric luminosities of 1.5\,$\times$\,10$^{39}$\,erg\,s$^{-1}$ for each star of the close binary system, and 3.0\,$\times$\,10$^{39}$\,erg\,s$^{-1}$ for the third component. This yields a total bolometric luminosity of about 6.0\,$\times$\,10$^{39}$\,erg\,s$^{-1}$. Following the empirical relation given by Bergh\"ofer et al. (\cite{BSDC}) applied to the three members of the multiple system, we infer expected individual X-ray luminosities and add them to obtain a total X-ray luminosity of about 1.0\,$\times$\,10$^{33}$\,erg\,s$^{-1}$ corresponding to an expected $L_\mathrm{X}$/$L_\mathrm{bol}$ ratio of about 1.7\,$\times$\,10$^{-7}$. Following the X-ray luminosities derived hereabove, the observed $L_\mathrm{X}$/$L_\mathrm{bol}$ ratio goes from 7.2\,$\times$\,10$^{-7}$ to 6.3\,$\times$\,10$^{-7}$ respectively for the April and the September {\it XMM-Newton} observations. These values yield X-ray luminosity excesses of about a factor 4.2 and 3.7 for both observations respectively. These values are significant considering the scatter of the relation of Bergh\"ofer et al. (\cite{BSDC}), suggesting that HD\,167971 is overluminous in X-rays.

\section{Comparison with previous observations \label{arch}}

HD\,167971 was observed with the {\it EINSTEIN} satellite on 31 March 1981 (sequence number 4240, 5.22\,ks). Chlebowski et al. (\cite{Chle}) reported an X-ray luminosity of 5.8\,$\times$\,10$^{33}$\,erg\,s$^{-1}$ for a 2\,kpc distance. This value is slightly larger than our {\it XMM-Newton} results. However, we compared the {\it EINSTEIN} results to ours following a more model independent way. We passed the model obtained in the case of the three thermal {\tt mekal} components through the {\it EINSTEIN} response matrix to obtain a simulated IPC spectrum. The model parameters are taken from the simultaneous EPIC-MOS fit for the April observation, and from EPIC-pn for the September one (see Table\,\ref{fitTTT}). Count rates were estimated between 0.2 and 3.5\,keV (the energy range of the IPC instrument) on the basis of these simulated spectra, and we obtained 0.049 $\pm$ 0.001\,cts\,s$^{-1}$ and 0.047 $\pm$ 0.001\,cts\,s$^{-1}$ respectively for the April and the September 2002 exposures. These values are in excellent agreement with the count rate given by Chlebowski et al. (\cite{Chle}), i.e. 0.050 $\pm$ 0.005\,cts\,s$^{-1}$. This result suggests that at the time of the {\it EINSTEIN} observation, HD\,167971 was in an emission state similar to our {\it XMM-Newton} observations.

\begin{figure}[t]
\begin{center}
\resizebox{8.5cm}{4.0cm}{\includegraphics{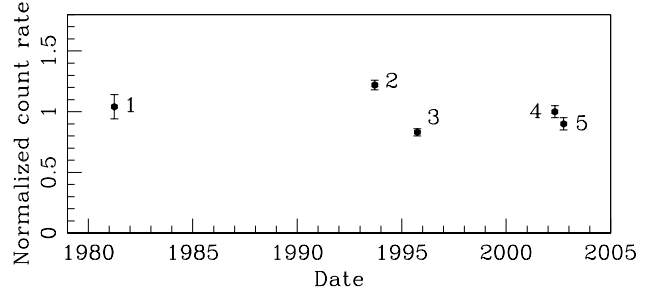}}
\caption{Normalized equivalent X-ray count rates of HD\,167971 arising from different observatories, as a function of time. 1: {\it EINSTEIN}-IPC, March 1981. 2: {\it ROSAT}-PSPC, September 1993. 3: {\it ROSAT}-HRI, September and October 1995. 4: {\it XMM-Newton}-EPIC, April 2002. 5: {\it XMM-Newton}-EPIC, September 2002. We note that the {\it ROSAT} and {\it EINSTEIN} points are averaged values from several observations spread over a few hours to a few days.\label{xarch}}
\end{center}
\end{figure}

This target was also observed with {\it ROSAT}. A PSPC observation (rp500298n00, 9.29 ks, performed between 13 September and 15 September 1993) reports a count rate of 0.129 $\pm$ 0.004\,cts\,s$^{-1}$. By folding the three-temperature model obtained for the fit of our EPIC data through the PSPC response matrix, we obtain 0.106 $\pm$ 0.002\,cts\,s$^{-1}$ and 0.103 $\pm$ 0.002\,cts\,s$^{-1}$ respectively for our two {\it XMM-Newton} observations between 0.1 and 2.5\,keV (the PSPC bandpass). These results suggest that HD\,167971 was in a higher emission state at the time of the {\it ROSAT} observation. The same procedure was applied in the case of a HRI observation performed between 12 September and 10 October 1995 (rh201995n00, 36.3 ks). The reported HRI count rate is 0.035 $\pm$ 0.001\,cts\,s$^{-1}$. The results we obtain by passing the model through the HRI response matrix are 0.042 $\pm$ 0.001\,cts\,s$^{-1}$ and 0.041 $\pm$ 0.001\,cts\,s$^{-1}$ respectively for the April 2002 and the September 2002 {\it XMM-Newton} observations, suggesting that at the moment of the HRI observation, HD\,167971 was in an emission state somewhat lower than that observed with {\it XMM-Newton}.

In summary, these results from archive data and from our {\it XMM-Newton} observations are plotted in Fig.\,\ref{xarch}. We note that we do not expect any confusion with other X-ray sources to affect the archive count rates quoted in this section. The count rates are normalized relative to the {\it XMM-Newton} count rate of April 2002 arbitrarily set to unity. The normalized count rate of the September 2002 {\it XMM-Newton} observation is determined following the flux ratio observed between 0.4 and 10.0 keV for the three-temperature model. The error bars on the {\it XMM-Newton} count rates are set to about 5\,\% in Fig.\,\ref{arch}. We consider that this value is more realistic than the standard deviations quoted in Table\,\ref{lum2}. We see that the X-ray flux of HD\,167971 undergoes a significant variability with a peak to peak amplitude of about 40 \%. However,  we note that this apparent variability relies mostly on the position of the {\it ROSAT}-PSPC point in Fig.\,\ref{xarch}, as other count rates suggest a more or less constant level. Moreover, we want to draw attention to the fact that the overall shape of the light curve plotted in Fig.\,\ref{xarch} is very similar to that presented in Fig.\,9 of De Becker et al.\,(\cite{paper1}) for HD\,168112, even if the amplitude of the apparent variability was larger in that case. Consequently this apparent variability might be attributed to some systematic effects affecting the various observations (common to HD\,168112 and HD\,167971) mentioned in this section. A detailed description of this behaviour requires a better data coverage than that available at the time of this study.

\section{Discussion \label{sect_disc}}
\subsection{Search for a colliding-wind signature}
As HD\,167971 is a multiple system whose components are O-stars, a wind-wind interaction is expected. Do our {\it XMM-Newton} data reveal the signature of colliding winds? Several questions should be considered to address this point:
\begin{enumerate}
\item[-]{\it Plasma temperature.} The results of the fittings with the three-temperature model point to a plasma temperature for the hardest component of about 2 -- 4\,keV, i.e. about 2.3\,$\times$\,10$^{7}$ to 4.6\,$\times$\,10$^{7}$\,K. Such high temperatures are usually not expected for the emission from shocks due to intrinsic wind instabilities, but are expected for shocks between the winds within a binary system (see e.g. Stevens et al.\,\cite{SBP}). These post-shock temperatures correspond to pre-shock velocities of about 1300 -- 1800 km\,s$^{-1}$, typical for winds colliding at speeds near the terminal velocity, as should be expected for long period wide binaries (Pittard \& Stevens \cite{PS}). For short period binary systems like the O5-8V + O5-8V pair in HD\,167971 (P $\sim$ 3.32 d), the winds have not reached their terminal velocities before they collide, and the temperatures are not expected to be so high. Such systems yield typical temperatures of at most 1\,$\times$\,10$^{7}$\,K. For instance, HD\,159176 (O7V + O7V) with a period of 3.367 d displays temperatures of about 2\,$\times$\,10$^{6}$ to 6\,$\times$\,10$^{6}$\,K (De Becker et al.\,\cite{DeB}), whilst the O7.5(f)III + O7.5(f)III close binary HD\,152248 (5.816\,d period) has plasma temperatures lower than 1\,$\times$\,10$^{7}$\,K (Sana et al.\,\cite{sana}). So, if the high temperatures we derived from our spectral fittings have a physical meaning, they are most probably related to the interaction between the wind of the O8I component and the combined winds of the close O5-8V + O5-8V binary system. We note however that high plasma temperatures can arise in the winds of individual stars if the magnetic field of the star is able to confine the stellar wind near the magnetic equatorial plane (Babel \& Montmerle \cite{BM}).
\item[-]{\it X-ray luminosity.} The luminosities we derived from our spectral fittings point to an excess attributable to a wind-wind interaction. This is compatible with a scenario where the colliding winds bring a significant contribution to the overall X-ray luminosity, in addition to the emission from the intrinsic shocks of the individual winds of the components of HD\,167971.
\item[-]{\it X-ray variability.} As discussed in Section \ref{var}, the question of the variability of the X-ray flux is rather complex. We do not detect any strong variability in the X-ray flux between our two {\it XMM-Newton} observations. Only a slight decrease possibly due to the eclipse of the O5-8V + O5-8V close binary system is observed. However, our discussion of archive data reveals that HD\,167971 possibly shows some variability, even if the data coverage is insufficient to constrain the time scale of this variability. If not due to some undetermined systematic effect (see last paragraph of Sect.\,\ref{arch}), the amplitude of this apparent variability seems to be rather high ($\sim$\,40\,\%). Even if there is no clear evidence that the three stars are physically related, this variability might be compatible with what could be expected from a long period and eccentric binary system. The count rates from {\it ROSAT} and {\it EINSTEIN} result indeed from a combination of several obervations spanning different phases of the orbit of the O5-8V + O5-8V binary system, and consequently are phase averaged values. Moreover, most of the count rates quoted in Fig.\,\ref{xarch} point to a low level, although it is not reasonable to think that all the pointings except that of {\it ROSAT}-PSPC fall at the moment of an eclipse. For these reasons, we do not expect the eclipse of the close binary system to be responsible for the variability illustrated in Fig.\,\ref{xarch}. However, considering the lack of reliability of this putative long term variability, we decided to attribute only little weight to this argument.
\end{enumerate}

Regarding these results, we realize that we find a probable signature of a wind-wind collision. Some clues, like the high plasma temperature in the context of the thermal model, point to a possible detection of the interaction between the close system and the third more luminous companion. With the large separations characterizing this interaction, the shocks of the collision zone are expected to be strongly adiabatic, resulting in a phase-locked X-ray variability scaling with 1/D, with D being the distance from one of the two components to the interaction zone (see Stevens et al.\,\cite{SBP}; Pittard \& Stevens\,\cite{PS}). According to this scenario, the X-ray emission should peak close to the periastron passage. Such a high temperature for the hard component of the spectrum, if it is thermal, was also reported in the case of HD\,168112 which is possibly a wide and eccentric binary system (De Becker et al.\,\cite{paper1}; Blomme et al.\,\cite{radio}). In the case of HD\,167971, the X-ray luminosity excess attributed to the wind collision is much more significant.

Quantifying the respective contributions from the two expected collision zones is not an easy task. To address such an issue, an accurate knowledge of the stellar properties of the three stars along with a better idea of their relative positions is needed. We can however tentatively adopt the following semi-quantitative approach. From Sect.\,\ref{ol}, we can assume that the cumulated intrinsic contributions from the three stars is about 10$^{33}$\,erg\,s$^{-1}$. As the overluminosity factor is about 4, the X-rays arising from the collision zone(s) amount to about 3\,$\times$\,10$^{33}$\,erg\,s$^{-1}$. On the other hand, for the O7V + O7V close binary system HD\,159176 (De Becker et al.\,\cite{DeB}) rather similar to the eclipsing binary harboured by HD\,167971\footnote{The bolometric luminosity provided by van Genderen et al.\,(\cite{vanG}) for the latter stars is intermediate between those given by Howarth \& Prinja\,(\cite{HP}) for the O5.5V and O6V spectral types.}, the same considerations lead to an X-ray luminosity for the collision zone of about 1.3\,$\times$\,10$^{33}$\,erg\,s$^{-1}$. If the O5-8V + O5-8V binary within HD\,167971 produces the same amount of X-rays, we estimate that the putative collision zone due to the third companion may produce more than twice as much X-rays as that of the close eclipsing binary. We mention however that the highest temperature component of the thermal model used to fit the EPIC spectra accounts for only about 10\,\% of this quantity, suggesting that the softest components account significantly for the putative wind-wind collision with the third star.

\subsection{Non-thermal emission from HD\,167971}
In the radio domain, the non-thermal nature of the emission from HD\,167971 is well established (Bieging et al.\,\cite{BAC}; Blomme et al.\,\cite{Blom2}). The high non-thermal radio flux suggests that it comes from the interaction between the close eclipsing binary and the third star. However, in the X-ray domain, the nature of the hard component of the spectrum is still unclear. As discussed in Sect.\,\ref{specan}, the pure thermal model and that including the power law yield results of similar quality. This ambiguity is a common feature in the study of X-ray spectra of massive stars (see e.g. De Becker \cite{DeBMT}; Rauw et al.\,\cite{9sgr}; De Becker et al.\,\cite{paper1}). If we make the assumption that a non-thermal emission component is responsible for the hard part of the spectrum of HD\,167971, we obtain photon index values close to 3, which is rather steep as compared to the value, i.e. 1.5, considered by Chen \& White (\cite{CW}) for X-rays produced through Inverse Compton scattering from relativistic electrons accelerated in strong shocks. For a discussion of such high values of the photon index, we refer to De Becker et al. (\cite{paper1}).

\begin{table*}
\caption{Comparison of the cases of 9\,Sgr (Rauw et al.\,\cite{sana}), HD\,168112 (De Becker et al.\,\cite{paper1}), HD\,167971 (this study), and HD\,159176 (De Becker et al.\,\cite{DeB}). The question marks for the multiplicity and the period stem from the fact that these are not well established.\label{compar}}
\begin{center}
\begin{tabular}{lcccc}
\hline
\hline
\vspace*{-0.3cm}\\
 & 9\,Sgr & HD\,168112 & HD\,167971 & HD\,159176 \\
\vspace*{-0.3cm}\\
\hline
\vspace*{-0.3cm}\\
Multiplicity & binary ? & binary ? & triple & binary \\
Period  & long period  ? & long period (1.4\,yr) ? & 3.3213 d & 3.367 d \\
  &  &  & + long period & \\
Thermal X-rays & soft + hard & soft + hard & soft + hard & soft \\
Hard X-rays & ambiguous & most probably & ambiguous & possible power \\
  &  & thermal &  & law tail \\
Synchrotron radio emission & yes & yes & yes & no \\
\vspace*{-0.3cm}\\
\hline
\end{tabular}
\end{center}
\end{table*}

\subsection{Comparison with other early-type stars \label{comp}}
Up to now, our campaign devoted to non-thermal radio emitters includes the observations of three targets: 9\,Sgr (Rauw et al.\,\cite{sana}), HD\,168112 (De Becker et al.\,\cite{paper1}), and HD\,167971 (this study). The next targets of this campaign are the non-thermal radio emitters of the Cyg\,OB2 association (\#8A, \#9 and \#5, see e.g. Waldron et al.\,\cite{wal}). Forthcoming studies based on {\it XMM-Newton} and {\it INTEGRAL} observations of these stars will bring new elements to the overall discussion of the high energy emission from massive stars. For instance, Cyg\,OB2\,\#8A is a well known non-thermal radio emitter that has recently been identified as a binary system by De Becker et al.\,(\cite{Let}), lending further support to the scenario where binarity is a necessary condition to observe non-thermal emission from massive stars. This idea is also supported by the results of Benaglia \& Koribalski (\cite{BK}) who discuss the case of four southern non-thermal radio emitters, among which three are confirmed binary systems. The need for a binary scenario to explain the non-thermal radio emission has independently been demonstrated by Van Loo (\cite{vlsac}) following a theoretical approach. Several works has recently been devoted to the non-thermal radio emission from massive binaries (Dougherty et al.\,\cite{DPal}; Pittard et al.\,\cite{PDC}). These recent models take into account several physical effects likely to affect the observed synchrotron spectrum (Razin effect, free-free absorption, IC cooling). Future developments of these models are expected to lead to an estimate of the non-thermal high-energy emission from massive binaries.

At this stage, it could be interesting to confront the results obtained in the case of the first three targets studied in this campaign. The summary of the results arising from these studies is provided in Table\,\ref{compar}. For these three objects, our analyses did not reveal any unambiguous power-law emission component in the X-ray spectra. A striking characteristic of their X-ray spectra is that the fit with thermal models reveals a rather hard emission, with plasma temperatures of a few 10$^{7}$\,K. HD\,167971 is a known multiple system and such high plasma temperature could be explained by a wind-wind interaction. HD\,168112 is not known to be a binary system, but was proposed to be a binary candidate by De Becker et al. (\cite{paper1}). The results of the study of Blomme et al. (\cite{radio}) lend further support to this idea, suggesting a period of about 1.4 yr based on archival radio and X-ray data, even if no unambiguous evidence of binarity has been found in the optical domain. In the case of 9\,Sgr, low amplitude radial velocity variations in the optical spectrum suggest it is a binary with a yet undetermined period (Rauw et al.\,\cite{9sgr}). If the high plasma temperatures observed for these three stars are indeed due to wind collisions within binary systems, the observed post-shock temperatures are related to rather high pre-shock wind velocities (typically at least $\sim$\,1000\,km\,s$^{-1}$). Such high velocities can only be achieved in binary systems where the winds collide after they have reached their terminal velocity, i.e. in systems with periods exceeding several days. In this scenario, the non-thermal radio emission is produced by a population of relativistic electrons accelerated in the shock due to the same wind-wind interaction. The putative non-thermal X-ray counterpart to this non-thermal radio emission has little chance to be detected, as it would probably be overwhelmed by the hard thermal component produced by the wind collision. If on the contrary the period were shorter, the pre-shock velocities would be lower and the X-ray thermal emission would be softer, allowing consequently the putative non-thermal X-ray emission to be detected in the hard part of the spectrum. However, the shorter orbital period required to possibly observe the non-thermal X-ray emission could cause the wind collision zone to be deeply embedded in the stellar wind material, thereby leading to a severe absorption of the synchrotron radio emission.

The fact that the conditions for detecting a non-thermal emission in the radio domain may be different from those for such a detection in the X-rays raises an interesting question. We may indeed wonder whether non-thermal radio emitters are ideal candidates to search for a non-thermal X-ray counterpart. To address this issue, let us consider some results obtained for other massive stars recently observed in the X-ray domain.

In the case of the short period massive binary HD\,159176 (OV7 + OV7), a high energy tail is observed in the X-ray spectrum. It can not be fitted with a thermal model\footnote{Optically thin plasma model (Mewe et al.\,\cite{mewe}; Kaastra \cite{ka}) or colliding wind model (Antokhin et al.\,\cite{ant}).}, but can be approximated with a power law (De Becker et al.\,\cite{DeB}). For that system, the typical temperature of the X-ray emitting plasma is sufficiently low to prevent the putative power law tail to be overwhelmed by the thermal X-rays from the colliding wind zone. This is due to the fact that the winds collide as they have not reached their terminal velocities. The non-detection of HD\,159176 (Bieging et al.\,\cite{BAC}) in the radio domain is compatible with the rather low flux expected for purely thermal radio emitters, and can be explained by the fact that any putative synchrotron emitting zone would be deeply embedded within the stellar winds. We mention also the case of WR\,110 (HD\,165688) whose binarity has not yet been established. Skinner et al. (\cite{skin}) reported a soft thermal X-ray spectrum, along with a high energy power law tail, whilst its radio spectrum is purely thermal.

Considering these recent results, we propose that the non-thermal radio emission could ideally be detected in the case of binaries with periods larger than several weeks. For instance, the shortest period well-established non-thermal radio emitter is Cyg\,OB2\,\#8A, with an orbital period of about 21.9\,d (De Becker et al.\,\cite{Let}). On the other hand, short period binaries are not expected to display a non-thermal radio emission because of the strong absorption by the wind material. However, their rather soft thermal X-ray emission could possibly unveil a power law component produced by inverse Compton scattering of UV photons. Moreover, the less diluted UV radiation field in the shock region of close binary systems is expected to lead to the production of a higher non-thermal X-ray flux, favoring its detection. In the context of this scenario, the simultaneous observation of non-thermal radiation in the X-ray (below 10.0\,keV) and radio domains appears rather unlikely.

\section{Other X-ray sources in the field of view\label{ngc}}
\subsection{Source list \label{list}}

\begin{figure*}[ht]
\begin{center}
\resizebox{17cm}{13cm}{\includegraphics{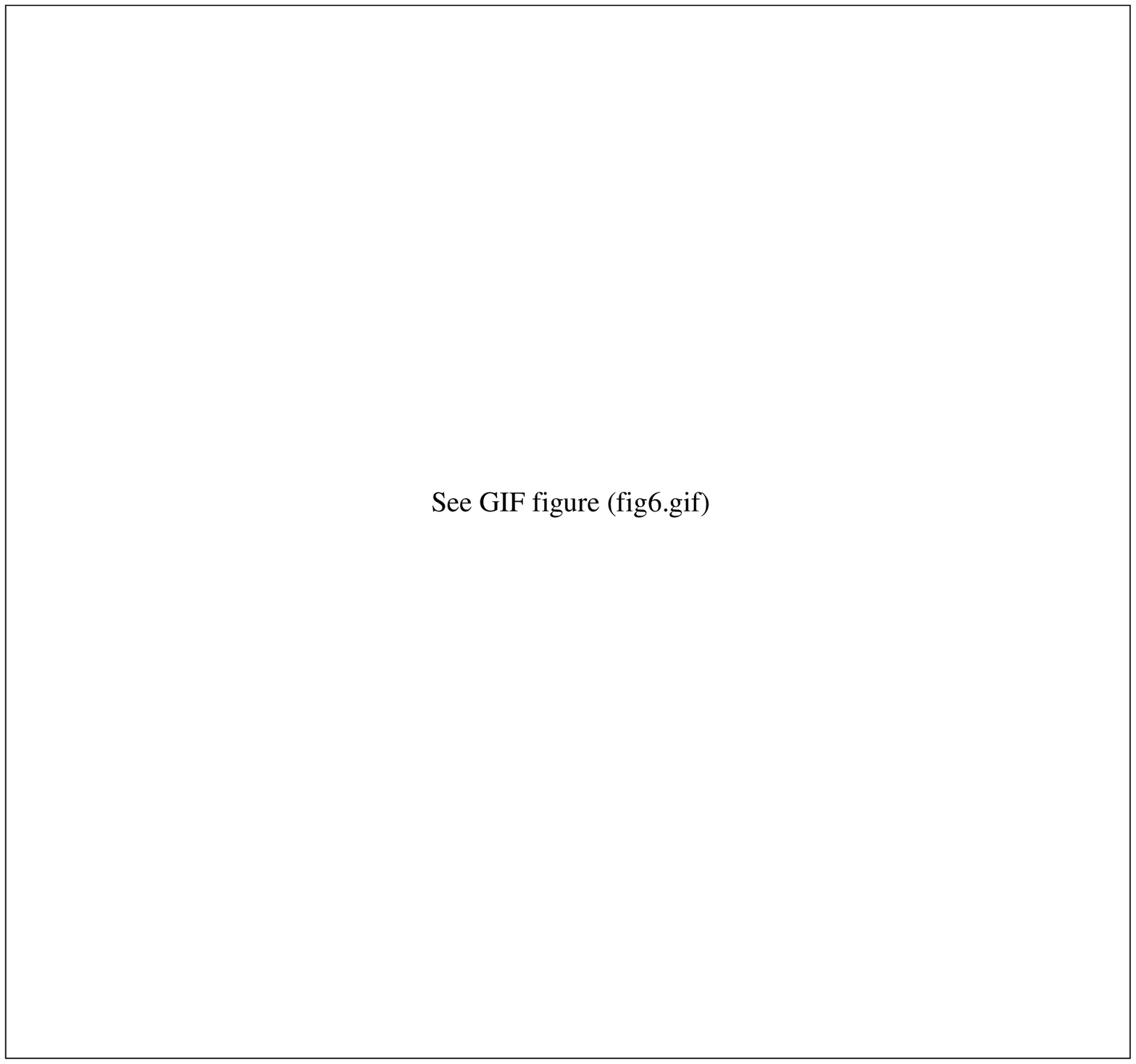}}
\caption{Image of the NGC\,6604 open cluster from the combined EPIC data sets of the two observations. The X-ray sources listed in Table\,\ref{xsrc} are labelled. The pixel size is 2.5\,$\arcsec$. The right ascension is increasing from the right to the left, and the declination is increasing from the bottom to the top. The width of the field is about 30\,$\arcmin$.\label{field}}
\end{center}
\end{figure*}

Beside HD\,167971 discussed in the previous sections, and HD168112 (De Becker et al.\,\cite{paper1}), other fainter X-ray sources are observed in the field of the EPIC cameras. To investigate the X-ray emission from these sources, we processed our data with the version 6.0.0 of the Science Analysis Software (SAS) to benefit from the latest version of the {\it edetect\_chain} metatask. We filtered the event lists of the April observation to reject the end of the exposure contaminated by a soft proton flare (see Sect.\,3.1 of De Becker et al.\,\cite{paper1}), thereby reducing the effective exposure time to about 9 and 6\,ks for EPIC-MOS and EPIC-pn respectively. Adding together the two observations and considering that EPIC-pn is about twice as sensitive as EPIC-MOS, we obtain an EPIC-MOS equivalent combined exposure time of about 80\,ks on NGC\,6604.

We first applied the source detection simultaneously to the data from the three EPIC instruments of our two observations to improve the detection efficiency. We used three energy bands respectively refered to as {\it S} (0.5--1.0\,keV), {\it M} (1.0-2.5\,keV), and {\it H} (2.5-10.0\,keV). The images in these three energy bands were extracted on the basis of the merged event lists of the two observations, for the three EPIC instruments respectively. We obtained a first source list using a detection likelihood threshold of 40. We then inspected each source of the list by eye to reject false detections due to background fluctuations and instrumental artifacts following the same approach as Rauw et al.\,(\cite{ngc6383}) and Naz\'e et al.\,(\cite{108}). Adopting the formalism described by Sana et al.\,(\cite{ngc6231}), we determined the likelihood thresholds adequate for individual datasets. The likelihood thresholds we finally adopted are 40 for the combined EPIC data sets, 40 for combined EPIC-MOS, 10 for individual EPIC-MOS, and 10 for EPIC-pn. After applying the same procedure to data sets resulting from individual instruments and/or observations, we added a few more sources to the list obtained for the full combined data set. We finally obtain a catalogue of 31 X-ray sources presented in Table\,\ref{xsrc} by order of increasing right ascension. The position of these sources in the EPIC field is shown in Fig.\,\ref{field}. All sources are identified following the naming conventions recommended by the {\it XMM SOC} and the IAU.

\begin{sidewaystable*}
\begin{center}
\caption{Characteristics of the X-ray sources in the field of view. The 31 sources are sorted by order of increasing right ascension. Source \#1 was identified as an extended source, and the 30 next ones as point sources. The count rates corrected for the exposure map are quoted for all three EPIC instruments. The missing values are due to a strong deviation from those of the other instruments because of CCD gaps. The hardness ratios are given for EPIC-pn only. The quoted counterparts are located within 4.5$\arcsec$ of the X-ray sources. Nr is the number of counterparts within the correlation radius, and $d$ is the angular separation between the X-ray source and its nearest counterpart. The error bars on the count rate represent the $\pm$\,1\,$\sigma$ Poissonian standard deviation.
 \label{xsrc}}\medskip
\tiny
\begin{tabular}{l l | c c c c c | c c c c c c | c c c c c | c c c } 
\hline\hline
\vspace*{-0.2cm}\\
\#& XMMU & MOS1 CR & MOS2 CR & pn CR & $HR_1$ & $HR_2$ & \multicolumn{6}{c|}{GSC 2.2} & \multicolumn{5}{c}{2MASS}& \multicolumn{3}{|c}{USNO B1.0}\\ 
& & 10$^{-3}$ cts\,s$^{-1}$ & 10$^{-3}$ cts\,s$^{-1}$ & 10$^{-3}$ cts\,s$^{-1}$ & & & Nr & $d$(") & Cat. id. & $R$ & $B$ & $V$ & Nr & $d$(") & $J$ & $H$ & $K_S$ & Nr & $d$(") & Cat. id. \\
(1) & (2) & (3) & (4) & (5) & (6) & (7) & (8) & (9) & (10) & (11) & (12) & (13) & (14) & (15) & (16) & (17) & (18) & (19) & (20) & (21) \\
\vspace*{-0.2cm}\\
\hline
\vspace*{-0.2cm}\\
1& J181731.4--120622.2$^a$  &  &  & 311.0$\pm$15.3 & 0.34$\pm$0.05 & --0.67$\pm$0.05 & 1 & 4.4 & S3001223315 & & 10.6 & 9.7 & 1 & 4.5 & 7.3 & 7.1 & 6.9 & 1 & 4.4 & 0778-0542759 \\
2& J181746.1--120542.3  & 6.1$\pm$1.0 & 5.8$\pm$1.0 & 13.1$\pm$1.9 & 0.55$\pm$0.14 & --0.17$\pm$0.16 & 1 & 2.0 & S300122014588 & 17.9 & & & 2 & 1.9 & 15.0 & 13.9 & 12.6 & 1 & 2.3 & 0779-0530036 \\
  & & & & & & & & & & & & & & 2.8 & 14.7 & 13.8 & 12.2 & & & \\
3& J181750.7--120400.4  & 4.4$\pm$1.0 & 2.6$\pm$0.9 & 12.1$\pm$1.9 & 0.51$\pm$0.14 & --0.62$\pm$0.20 & 1 & 2.2 & S3001220900 & 13.0 & 14.4 & 13.3 & 1 & 1.5 & 11.4 & 11.1 & 10.9 & 1 & 1.9 & 0779-0530128 \\
4& J181750.7--120506.3  & 3.5$\pm$0.8 & 4.2$\pm$0.8$^*$ & 11.8$\pm$1.8 & 0.30$\pm$0.14 & --0.47$\pm$0.22 & 1 & 1.7 & S300122015025 & 15.7 & 18.7 & 16.3 & 1 & 1.4 & 13.4 & 12.6 & 12.4 & 1 & 2.3 & 0779-0530127 \\
5& J181806.0--121433.8$^b$  & 231.4$\pm$4.4$^*$ & 229.6$\pm$4.1$^*$ & 673.3$\pm$8.3$^*$ & 0.23$\pm$0.01$^*$ & --0.75$\pm$0.01$^*$ & 1 & 1.6 & S30012201153 & & 8.3 & 7.6 & 1 & 1.6 & 5.5 & 5.3 & 5.1 & 1 & 1.6 & 0777-0543906 \\
6& J181808.4--120824.9  & 1.3$\pm$0.4 & 1.2$\pm$0.4 & 5.7$\pm$0.9 & 0.51$\pm$0.31 & 0.42$\pm$0.15 & 1 & 2.6 & S300122012953 & 17.9 & & & 1 & 2.2 & 15.6 & 15.2 & 14.7 & 0 & & \\
7& J181808.6--120851.2  & 2.6$\pm$0.5 & 2.1$\pm$0.4 & 3.6$\pm$0.8$^*$ & 1.00$\pm$0.43$^*$ & 0.53$\pm$0.20$^*$ & 0 & & & & & & 1 & 2.6 & 15.2 & 13.8 & 13.4 & 1 & 3.2 & 0778-0543325 \\
8& J181810.0--121048.5$^c$  & 12.5$\pm$0.8 & 13.1$\pm$0.9 & 36.6$\pm$1.9$^*$ & 0.38$\pm$0.05$^*$ & --0.80$\pm$0.05$^*$ & 1 & 2.1 & S3001220165 & & 10.0 & 9.3 & 1 & 2.2 & 7.7 & 7.5 & 7.4 & 1 & 2.1 & 0778-0543346 \\
9& J181810.6--120409.3  & 2.1$\pm$0.5 & 2.8$\pm$0.6 & 7.7$\pm$1.2 & 0.75$\pm$0.16 & --0.16$\pm$0.17 & 1 & 1.8 & S300122015589 & 14.3 & 16.8 & 15.3 & 1 & 1.5 & 11.3 & 10.2 & 9.4 & 1 & 1.5 & 0779-0530455 \\
10& J181813.1--121503.1  & 4.6$\pm$0.7 & 4.2$\pm$0.7 & 3.8$\pm$0.9 & 0.89$\pm$0.16 & --0.82$\pm$0.23 & 1 & 4.4 & S30012209459 & 17.5 & & & 1 & 3.3 & 15.3 & 14.4 & 14.3 & 1 & 3.5 & 0777-0544085 \\
11& J181815.3--121148.7  & 4.6$\pm$0.6 & 4.5$\pm$0.6 & 13.4$\pm$1.2 & 0.72$\pm$0.08 & --0.13$\pm$0.09 & 1 & 1.2 & S300122011032 & 18.5 & & & 1 & 1.3 & 15.2 & 14.2 & 14.0 & 0 & & \\
12& J181815.7--120842.8  & 1.2$\pm$0.4 & 1.5$\pm$0.4 & 2.6$\pm$0.7 & 1.00$\pm$0.31 & 0.34$\pm$0.24 & 0 & & & & & & 0 & & & & & 0 & & \\
13& J181816.9--121617.4  & 3.3$\pm$0.6 & 1.2$\pm$0.5 & 3.7$\pm$0.9$^*$ & 1.00$\pm$0.39$^*$ & 0.08$\pm$0.27$^*$ & 0 & & & & & & 0 & & & & & 0 & & \\
14& J181819.3--121616.9  & 5.4$\pm$0.8 & 3.6$\pm$0.6 & 13.3$\pm$1.7$^*$ & 0.54$\pm$0.12$^*$ & --0.41$\pm$0.14$^*$ & 0 & & & & & & 0 & & & & & 0 & & \\
15& J181820.5--121736.3  & 1.7$\pm$0.5 & 0.6$\pm$0.3 & 4.2$\pm$0.9 & 1.00$\pm$0.12 & 0.16$\pm$0.21 & 0 & & & & & & 0 & & & & & 1 & 1.1 & 0777-0544362 \\
16& J181822.6--121503.4  & 7.3$\pm$0.8 & 6.6$\pm$0.7 & 15.2$\pm$1.5$^*$ & 0.85$\pm$0.17$^*$ & 0.45$\pm$0.08$^*$ & 0 & & & & & & 0 & & & & & 0 & & \\
17& J181823.2--120719.4  & 1.1$\pm$0.4 & 2.2$\pm$0.4 & 4.1$\pm$0.8 & 0.38$\pm$0.33 & 0.42$\pm$0.18 & 0 & & & & & & 0 & & & & & 1 & 2.5 & 0778-0543806 \\
18& J181827.3--120613.4  & 3.0$\pm$0.5 & 3.5$\pm$0.5$^*$ & 11.9$\pm$1.2 & 0.45$\pm$0.09 & --0.67$\pm$0.11 & 1 & 0.9 & S300122014289 & 18.1 & & & 1 & 0.1 & 15.2 & 14.1 & 13.8 & 1 & 0.2 & 0778-0544019 \\
19& J181830.3--121359.1  & 4.1$\pm$0.5 & 3.5$\pm$0.5 & 10.4$\pm$1.0 & 0.61$\pm$0.08 & --0.37$\pm$0.11 & 1 & 1.6 & S30012209935 & 17.2 & & & 1 & 1.6 & 15.2 & 14.3 & 14.0 & 1 & 1.6 & 0777-0545005 \\
20& J181832.0--121740.3  & 1.6$\pm$0.5 & 1.4$\pm$0.5$^*$ & 4.3$\pm$1.0 & 0.60$\pm$0.20 & --0.60$\pm$0.27 & 1 & 2.4 & S30012208102 & 15.2 & 18.9 & 17.7 & 2 & 1.4 & 13.8 & 12.9 & 13.3 & 2 & 2.3 & 0777-0545104 \\
  & & & & & & & & & & & & & & 2.8 & 13.5 & 12.6 & 13.1 & & 2.9 & 0777-0545111 \\
21& J181832.1--121605.0  & 3.5$\pm$0.6 & 2.4$\pm$0.6$^*$ & 7.6$\pm$1.1 & 0.54$\pm$0.12 & --0.51$\pm$0.16 & 1 & 3.5 & S300122028708 & 18.4 & & & 2 & 2.6 & 14.7 & 13.8 & 13.5 & 1 & 3.3 & 0777-0545106 \\
  & & & & & & & & & & & & & & 4.0 & 14.2 & 13.2 & 12.8 & & & \\
22& J181837.1--120601.7  & 1.6$\pm$0.4 & 1.3$\pm$0.4 & 8.1$\pm$1.0$^*$ & 0.09$\pm$0.12$^*$ & --1.00$\pm$0.10$^*$ & 1 & 2.6 & S300122014474 & 14.1 & 16.2 & 15.6 & 1 & 1.5 & 13.5 & 12.9 & 12.9 & 1 & 2.8 & 0779-0531112 \\
23& J181839.5--120935.3  & 1.3$\pm$0.3 & 1.3$\pm$3.3 & 5.2$\pm$0.9$^*$ & 0.60$\pm$0.16$^*$ & --0.30$\pm$0.19$^*$ & 1 & 3.8 & S30012201041 & 13.5 & 14.5 & 13.8 & 2 & 1.6 & 14.2 & 13.0 & 12.3 & 1 & 3.5 & 0778-0544645 \\
  & & & & & & & & & & & & & & 4.0 & 12.5 & 12.2 & 12.1 & & & \\
24& J181840.9--120623.7$^d$  & 55.3$\pm$1.8 & 59.2$\pm$1.9 &  &  &  & 1 & 0.6 & S3001220959 & & 9.2 & 8.6 & 1 & 0.6 & 6.9 & 6.7 & 6.6 & 1 & 0.6 & 0778-0544725 \\
25& J181844.3--120752.3  & 12.1$\pm$0.9 & 13.5$\pm$0.9 & 45.6$\pm$1.9 & --0.37$\pm$0.04 & --0.89$\pm$0.06 & 0 & & & & & & 0 & & & & & 0 & & \\
26& J181901.2--120050.8  & 4.6$\pm$0.9 & 3.9$\pm$0.8 & 16.4$\pm$2.2 & 0.76$\pm$0.12 & --0.06$\pm$0.14 & 1 & 1.0 & S300122017698 & 15.2 & 18.5 & 16.6 & 1 & 1.2 & 12.0 & 11.0 & 10.5 & 1 & 1.0 & 0779-0532345 \\
27& J181903.0--120638.4  & 4.3$\pm$0.8 & 3.2$\pm$0.6 & 12.2$\pm$1.5 & --0.26$\pm$0.11 & --0.70$\pm$0.27 & 2 & 2.3 & S3001220971 & & 10.5 & 9.8 & 2 & 2.4 & 8.3 & 8.1 & 8.0 & 1 & 2.3 & 0778-0546021 \\
  &  &  &  &  &  &  &  & 3.4 & S300122014341 & 12.0 & & & & 4.2 & 13.8 & 12.9 & 12.7 & & & \\
28& J181904.9--120918.7$^e$  &  & 2.8$\pm$0.6 & 7.3$\pm$1.3$^*$ & 1.00$\pm$0.18$^*$ & 0.54$\pm$0.13$^*$ & 0 & & & & & & 1 & 2.0 & 14.2 & 13.1 & 12.8 & 0 & & \\
29& J181905.2--120306.5  & 2.9$\pm$0.7 & 5.1$\pm$0.9 & 9.1$\pm$1.7 & 0.99$\pm$0.09 & --0.07$\pm$0.18 & 0 & & & & & & 2 & 0.9 & 15.0 & 14.0 & 13.5 & 2 & 1.1 & 0779-0532614 \\
  & & & & & & & & & & & & & & 3.8 & 15.3 & 14.3 & 14.0 & & 2.6 & 0779-0532625 \\
30& J181906.8--121018.2  & 5.4$\pm$0.9 & 7.0$\pm$0.9 & 18.0$\pm$1.9$^*$ & 0.60$\pm$0.19$^*$ & 0.60$\pm$0.08$^*$ & 0 & & & & & & 1 & 4.5 & 15.4 & 14.6 & 14.1 & 1 & 3.0 & 0778-0546278 \\
31& J181909.9--121753.4  & 5.7$\pm$1.1 & 8.2$\pm$1.3 & 12.8$\pm$2.1 & 1.00$\pm$0.05 & --0.23$\pm$0.17 & 1 & 1.0 & S300122028430 & 17.3 & & & 1 & 1.3 & 15.1 & 14.2 & 13.9 & 1 & 1.1 & 0777-0547481 \\
\hline
\end{tabular}
\end{center}
\small
$^a$ This X-ray source is associated to the O6V((f)) star HD\,167834 (BD\,-12$^\circ$\,4969). We quote the count rates obtained as an extended source detection. The X-ray properties quoted are from EPIC-pn data of the April observation only, as this source was not in the field of the other data sets.\\
$^b$ HD\,167971. The large difference between the count rates quoted for this star in this table and the observed ones given in Table\,\ref{lum2} are explained by the exposure map correction applied during the source detection procedure.\\
$^c$ BD\,-12$^\circ$\,4982 (O9.5I or III, see text).\\
$^d$ HD\,168112 : the EPIC-pn count rate is not quoted as a CCD gap crosses the sources region. The analysis described by De Becker et al.\,(\cite{paper1}) was performed with different screening criteria than for the current analysis to deal with the EPIC-pn data.\\
$^e$ We note that even if no optical counterpart was found in the catalogues used for the correlation, a star appears clearly on the DSS image centered on this position.\\ 
$^*$ This value is possibly affected by a CCD gap located at $<$10 arcsec, but is nevertheless quoted.
\end{sidewaystable*}
\normalsize

\subsection{Source identification}

The position of the 31 X-ray sources were cross-correlated with three catalogues: the Guide Star Catalogue (GSC, version 2.2)\footnote{The Guide Star Catalogue-II is a joint project of the Space Telescope Institute and the Osservatorio Astronomico di Torino.}, the Two Micron All Sky Survey (2MASS, Skrutskie et al.\,\cite{2mass}), and the US Naval Observatory (USNO-B1.0, Monet et al.\,\cite{usno}). We derived the optimal correlation radius following the procedure described by Jeffries et al.\,(\cite{jeff}). In this way, we find that a cross-correlation radius of 4.5\,arcsec includes the majority of the true correlations while rejecting most of the spurious correlations. With a radius of 4.5\,arcsec we expect to achieve at least 23 true and only about 2 spurious correlations. We also assume that the position of the source on the detector has little impact on the correlation radius. We should however keep in mind that this assumption is mostly valid close to the center of the field for a given instrument, but could possibly bias our source identification procedure for large off-axis angles.

\begin{figure*}[ht]
\begin{center}
\resizebox{17.5cm}{9.0cm}{\includegraphics{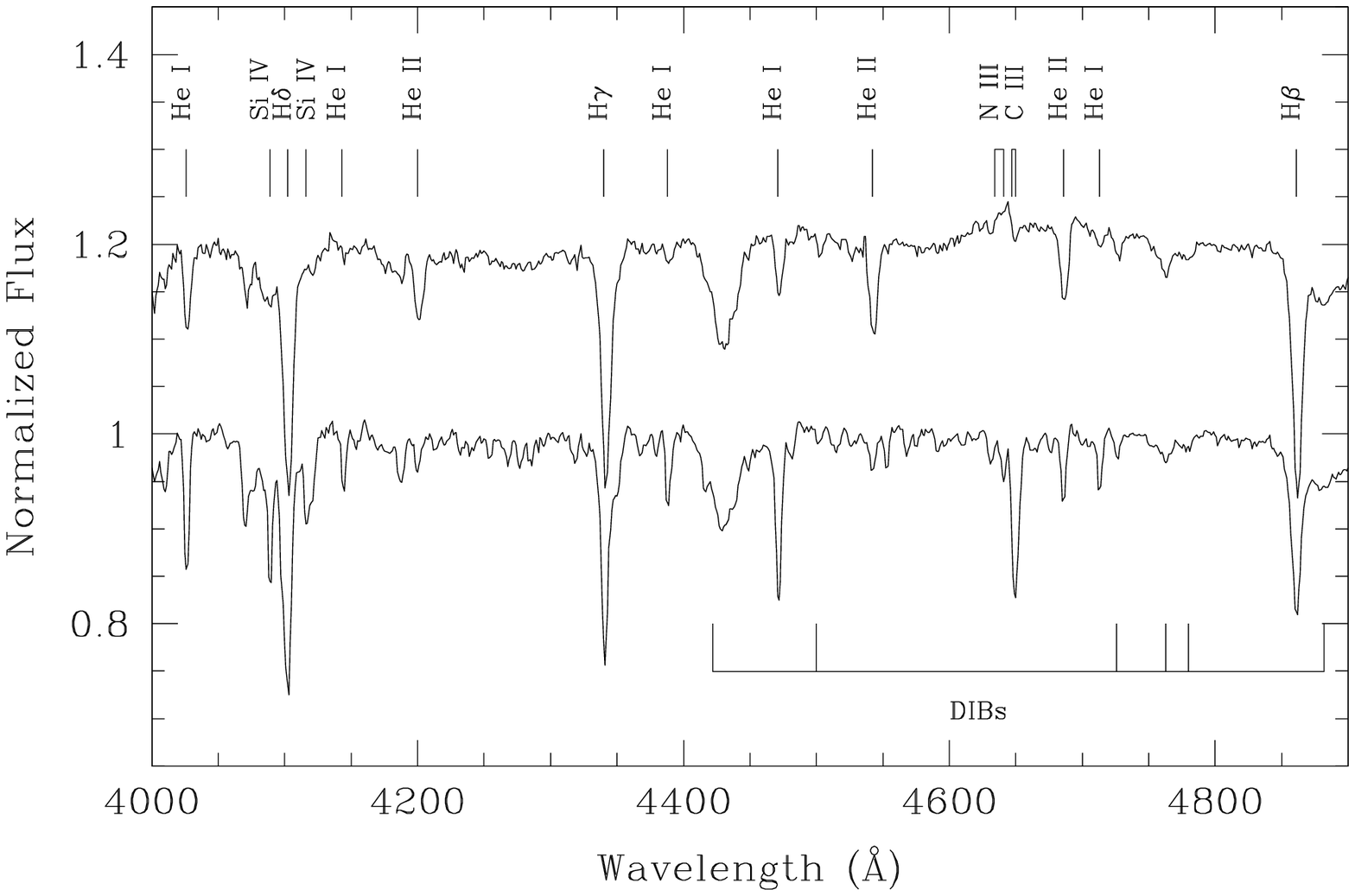}}
\caption{Optical normalized spectra obtained between 4000 and 4900\,\AA\, respectively for the sources \#1 (HD\,167834, {\it top spectrum}) and \#8 (BD\,-12$^\circ$\,4982, {\it bottom spectrum}). \label{opt}}
\end{center}
\end{figure*}

\begin{figure}
\begin{center}
\resizebox{8.2cm}{8cm}{\includegraphics{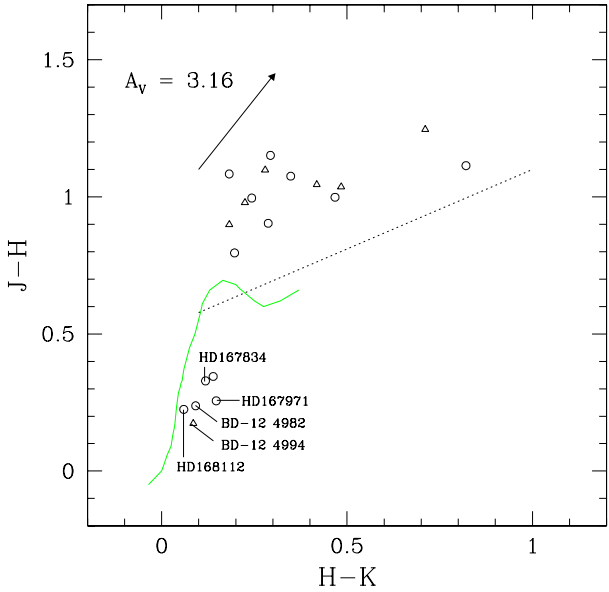}}
\end{center}
\caption{$JHK$ colour-colour diagram of the 2MASS counterparts of the X-ray sources in the EPIC field of view around NGC\,6604. The heavy solid line yields the intrinsic near-IR colours of main sequence stars following Bessell \& Brett (\cite{BessB}), whereas the reddening vector is illustrated for $A_V = 3.16$. Open circles and triangles stand for EPIC sources having respectivley a single or two 2MASS counterpart(s). The dotted straight line yields the locus of dereddened colours of classical T\,Tauri stars according to Meyer et al.\ (\cite{meyer}). \label{2mass}}
\end{figure}

Among the 31 X-ray sources quoted in Table\,\ref{xsrc}, five sources (\#12, \#13, \#14, \#16 and \#25) have neither a GSC, 2MASS nor USNO counterpart. 20 (respectively 23) have at least one optical counterpart in the GSC (resp. USNO) catalogue, and some sources have possibly two optical (\#20, \#27 and \#29) counterparts. No other positive correlations were found with the sources included in the photometric survey of Forbes \& DuPuy\,(\cite{FD}) within a 10\,arcsec radius. A total of 18 EPIC sources in the field of view of NGC\,6604 have a single 2MASS counterpart while another 6 sources have two infrared counterparts. In Fig.\,\ref{2mass}, we show the $JHK$ colour-colour diagram of those objects that have quality flags A or B for the measurements of all three individual near-IR magnitudes. We used the March 2003 update of the colour transformations, initially derived by Carpenter (\cite{Carpenter}) and available on the 2MASS website\footnote{\tt http://www.ipac.caltech.edu/2mass/index.html}, to convert the $J - H$ and $H - K_s$ colours to the homogenized photometric system introduced by Bessell \& Brett (\cite{BessB}). We thus exclude all objects that have colours and magnitudes that are either subject to large uncertainties or are only upper limits due to non-detections. Using the extinction law of Rieke \& Lebofsky (\cite{RL}) and assuming $R_V = A_V/E(B - V) = 3.1$ (Barbon et al.\ \cite{Bar}), we also show the reddening vector for $E(B - V) = 1.02$ as was found to be appropriate for NGC\,6604 (Barbon et al.\ \cite{Bar}). 

There are three broad groups that appear in Fig.\,\ref{2mass}. The first group consists of five sources that are clearly associated with early-type stars: \#1 (HD\,167834), \#5 (HD\,167971), \#8 (BD $-12^{\circ}$\,4982), \#24 (HD\,168112) and \#27 (BD $-12^{\circ}$\,4994). In the second group, we find objects around $H - K \sim 0.3$ and $J - H \sim 1.0$. These objects have colours that are consistent with slightly reddened late type (early G to late K) main-sequence or giant stars. Finally, a small group of objects (\#9, \#26 and one of the two counterparts of the three sources \#21, \#23 and \#29) have $H - K > 0.4$. Assuming that they are affected by the same reddening as NGC\,6608 ($E(B - V) = 1.02$), these objects would be associated with very late (mostly late M) main sequence stars. If these sources were normal main sequence stars belonging to NGC\,6604, it seems rather unlikely that we would be able to detect their X-ray emission. An alternative possibility could be that they are X-ray bright pre-main sequence stars with a moderate IR excess. In fact, their dereddened infrared colours are in broad agreement with the intrinsic colours of classical T\,Tauri stars (Meyer et al.\ \cite{meyer}).\\ 

Of course, a fraction of the X-ray selected objects might in fact be foreground stars or background sources. NGC\,6604 ($l_{\rm II} = 18.26^{\circ}$, $b_{\rm II} = 1.69^{\circ}$) lies very close to the Galactic plane and the total Galactic neutral hydrogen column density along this direction must therefore be quite large. This should produce a substantial absorption of X-ray photons from extragalactic background sources. Because of the Galactic coordinates of NGC\,6604, estimating this column density accurately is a very difficult task. For instance, the {\it DIRBE/IRAS} extinction maps provided by Schlegel et al.\ (\cite{schlegel}) yield a good estimate of the total Galactic $E(B - V)$ at Galactic latitudes above $|b_{\rm II}| \geq 5^{\circ}$, but are subject to very large uncertainties near the Galactic plane. With these limitations in mind, the maps of Schlegel et al.\ (\cite{schlegel}) suggest $E(B - V) \sim 3.5 \pm 0.5$ for NGC\,6604 (corresponding to $N_{\rm H} \sim 2 \times 10^{22}$\,cm$^{-2}$). 
Assuming that extragalactic background sources have a power-law spectrum with a photon index of 1.4, and are subject to a total interstellar absorption of $2 \times 10^{22}$\,cm$^{-2}$, the detection limits $\sim 4.0 \times 10^{-3}$ and $1.5 \times 10^{-3}$\,cts\,s$^{-1}$ for the pn and MOS detectors translate into unabsorbed fluxes of $2.1 \times 10^{-14}$\,erg\,cm$^{-2}$\,s$^{-1}$ and $6.0 \times 10^{-14}$\,erg\,cm$^{-2}$\,s$^{-1}$ in the 0.5 -- 2.0\,keV and 2.0 -- 10\,keV band respectively. From the $\log{N}$ -- $\log{S}$ relation of Giacconi et al.\ (\cite{giacconi}), we would expect to first order to detect around 8 -- 10 extragalactic sources over the EPIC field of view. We emphasize that these objects should be detected as rather hard sources. Hence, several of the hard sources quoted in Table\,\ref{xsrc} that lack an optical counterpart may actually be associated with AGN.

Optical spectra of the counterparts of EPIC sources \#1, \#8 and \#14 were obtained on June 22, 2004 with the EMMI instrument mounted on ESO's 3.5\,m New Technology Telescope (NTT) at La Silla. The EMMI instrument was used in the RILD low dispersion spectroscopic mode with grism \# 5 (600 grooves\,mm$^{-1}$) providing a wavelength coverage from about 3800 to 7020\,\AA\ with a spectral resolution of 5.0\,\AA\ ({\tt FWHM} of the He-Ar lines). The slit width was set to 1\,arcsec and the exposure times were 1\,min for the two bright sources \#1 and \#4, and 40\,min for the counterpart of source \#14. The data were reduced in the standard way using the {\tt long} context of the {\sc midas} package.

The optical spectrum of source \#1 (HD\,167834) is given in the top part of Fig.\,\ref{opt}. We obtain equivalent widths (EW) of 0.36 and 0.69\,\AA\, respectively for the \ion{He}{i} $\lambda$ 4471 and \ion{He}{ii} $\lambda$ 4542 lines. We estimate that the typical error on the estimate of EWs is about 5-10\,\%. According to the classification criterion given by Mathys (\cite{math}), and considering that the \ion{N}{iii} $\lambda$$\lambda$ 4634-41 lines are in emission and the \ion{He}{ii} $\lambda$ 4686 line is in strong absorption, we derive an O6V((f)) spectral type. We note also that the \ion{C}{iii} $\lambda$ 5696 line is in weak emission. In the case of source \#8 (BD\,-12$^\circ$\,4982, bottom spectrum of Fig.\,\ref{opt}), the ratio of the EW of the \ion{He}{i} $\lambda$ 4471 and \ion{He}{ii} $\lambda$ 4542 lines leads to an O9.5 spectral type. This spectral type is in excellent agreement with that given by Barbon et al.\,(\cite{Bar}, their source \#32). Following the EWs of the \ion{He}{i} $\lambda$ 4388 line (0.36\,\AA\,) and of the \ion{He}{ii} $\lambda$ 4686 line (0.31\,\AA\,), we derive a supergiant luminosity class, whilst the EWs of the \ion{Si}{iv} $\lambda$ 4088 line (0.54\,\AA\,) and of the \ion{He}{i} $\lambda$ 4143 line (0.29 \AA\,) point to a giant luminosity class. We note however that the spectrum of this star is very similar to that of the O9.7Iab star $\mu$\,Nor (HD\,149038) given by Walborn \& Fitzpatrick (\cite{WF}). Finally, for the source \#14, possibly associated to Cl*\,NGC\,6604\,FD\,61 (following the nomenclature proposed by Forbes \& DuPuy\,\cite{FD}) even if no optical counterpart is quoted in Table\,\ref{xsrc}, we derive a G6 spectral type, with an uncertainty of about 2 spectral types. We note that this optical counterpart is however located at about 6\,arcsec of the X-ray source \#14. Two possible 2MASS infrared counterparts are approximately located at the position, but the precision of their photometry is very poor.

\begin{table*}[ht]
\caption{Parameters for models fitted to the spectra of sources \#1 between 0.4 and 10.0\,keV, and \#8 and \#25 between 0.4 and 5.0\,keV. The data are from the EPIC-pn observation for source \#1, and from the three EPIC instruments for sources \#8 and \#25. The parameters have the same meaning as in Tables\,\ref{fitTTT} and \ref{fitTTP}. For sources \#1 and \#8, two absorption columns were used accounting respectively for the ISM and the local absorption (same ionized wind absorption model as for HD\,167971, left as a free parameter). The ISM columns were frozen at 0.73\,$\times$\,10$^{22}$ and 0.55\,$\times$\,10$^{22}$\,cm$^{-2}$ respectively for sources \#1 and \#8. In the case of source \#25, a unique absorption column (ISM + local) was used and left as a free parameter. The error bars represent the 90\% confidence interval. The upper and lower parts of the table are respectively devoted to the purely thermal models and to the model including a power law. The observed fluxes quoted in the last column are affected by large errors due to the large relative error on the normalization parameters (Norm$_1$ and Norm$_2$), and no significant variability between the April and September observations can be claimed on the basis of these values.\label{fitsrc}}
\begin{center}
\begin{tabular}{llccccccc}
\hline
\hline
\vspace*{-0.3cm}\\
Source & Obs. & $N_{\mathrm{H}}$ & $kT_1$ & Norm$_1$ & $kT_2$ & Norm$_2$ &  $\chi^2_\nu$ (d.o.f.) & Obs. Flux \\
 &	& (10$^{22}$\,cm$^{-2}$) & (keV) &  & (keV) &   &  & (erg\,cm$^{-2}$\,s$^{-1}$) \\
\hline
\vspace*{-0.3cm}\\
\#1 & April & 0.91$^{1.18}_{0.56}$ & 0.19$^{0.23}_{0.17}$ & 6.20$^{15.09}_{1.68}$\,$\times$\,10$^{-2}$ & 1.38$^{2.02}_{1.07}$ & 1.20$^{1.57}_{0.81}$\,$\times$\,10$^{-3}$ & 0.97 (78) & 8.0\,$\times$\,10$^{-13}$\\
\vspace*{-0.3cm}\\
\hline
\vspace*{-0.3cm}\\
\#8 & April & 0.71$^{1.06}_{0.43}$ & 0.71$^{0.86}_{0.62}$ & 0.26$^{4.75}_{2.73}$\,$\times$\,10$^{-3}$ & -- & -- & 0.73 (66) & 6.8\,$\times$\,10$^{-14}$\\
\vspace*{-0.3cm}\\
 & September & 0.69$^{0.86}_{0.49}$ & 0.71$^{0.80}_{0.64}$ & 0.30$^{3.84}_{2.12}$\,$\times$\,10$^{-3}$ & -- & -- & 0.85 (178) & 7.9\,$\times$\,10$^{-14}$\\
\vspace*{-0.3cm}\\
\hline
\vspace*{-0.3cm}\\
\#25 & April & 0.72$^{0.84}_{0.51}$ & 0.23$^{0.33}_{0.19}$ & 1.12$^{3.51}_{0.00}$\,$\times$\,10$^{-3}$ & -- & -- & 0.63 (95) & 7.9\,$\times$\,10$^{-14}$\\
\vspace*{-0.3cm}\\
 & September & 0.76$^{0.93}_{0.59}$ & 0.24$^{0.32}_{0.18}$ & 0.90$^{4.14}_{0.00}$\,$\times$\,10$^{-3}$ & -- & -- & 0.87 (133) & 6.1\,$\times$\,10$^{-14}$\\
\vspace*{-0.3cm}\\
\hline
\hline
\vspace*{-0.3cm}\\
 & 	& $N_{\mathrm{H}}$ & $kT$ & Norm$_1$ & $\Gamma$ & Norm$_2$ &  $\chi^2_\nu$ (d.o.f.) & Obs. Flux \\
 &	& (10$^{22}$\,cm$^{-2}$) & (keV) &  &  &  &  & (erg\,cm$^{-2}$\,s$^{-1}$) \\
\hline
\vspace*{-0.3cm}\\
\#1 & April & 0.20$^{1.24}_{0.00}$ & 0.58$^{0.66}_{0.46}$ & 1.43$^{2.61}_{0.47}$\,$\times$\,10$^{-3}$ & 2.98$^{7.28}_{1.19}$ & 4.04$^{12.96}_{2.31}$\,$\times$\,10$^{-4}$ & 1.16 (78) & 7.9\,$\times$\,10$^{-13}$\\
\vspace*{-0.3cm}\\
\hline
\end{tabular}
\end{center}
\end{table*}

\begin{figure}[h]
\begin{center}
\resizebox{8.5cm}{6.0cm}{\includegraphics{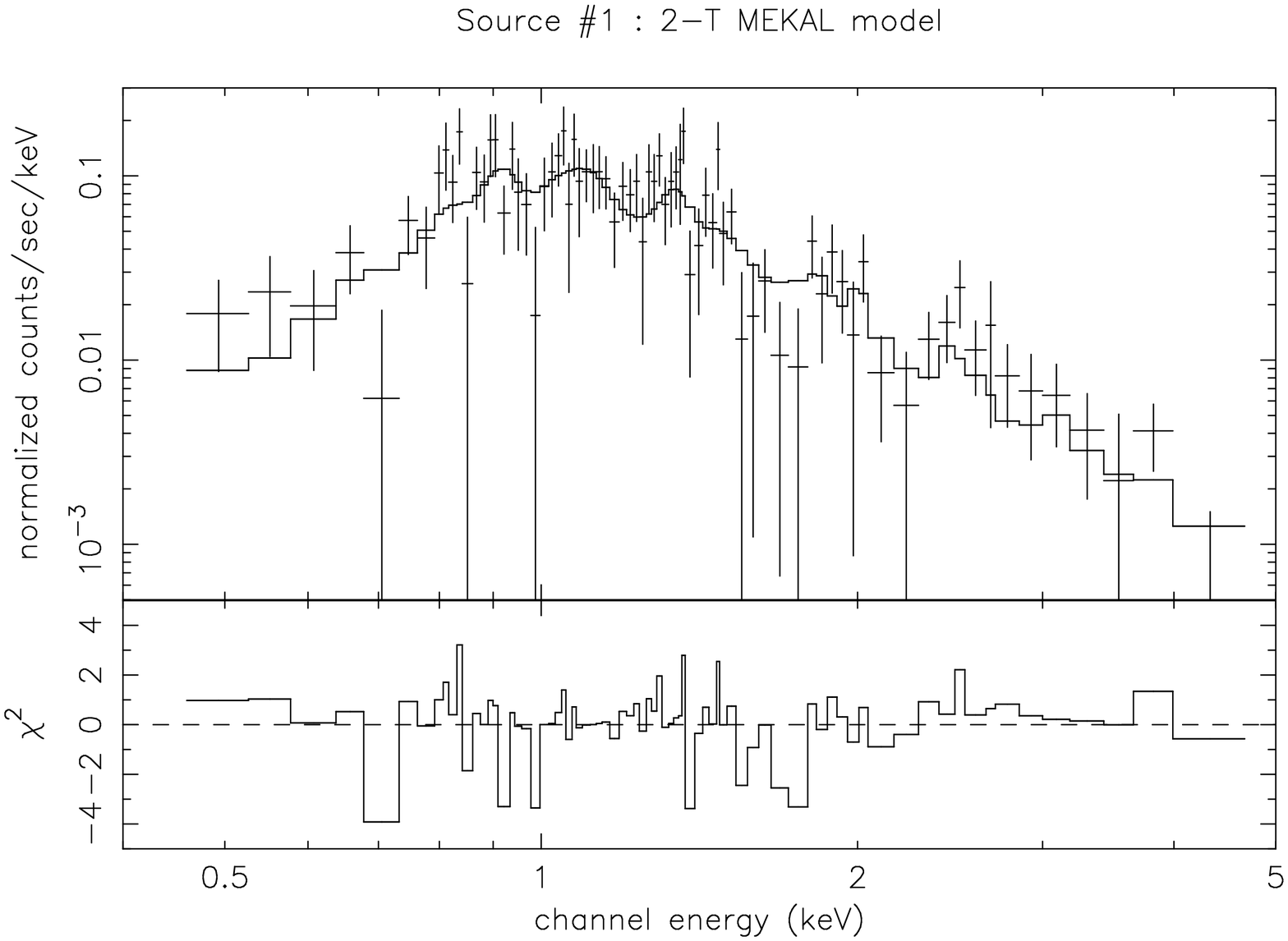}}
\caption{EPIC-pn spectrum of source \#1 (HD\,167834) obtained in April 2002, fitted with a {\tt wabs$_\mathrm{ISM}$*wind*(mekal+mekal)} model. The ISM absorption column was frozen at 0.73\,$\times$\,10$^{22}$\,cm$^{-2}$\label{sp1}}
\end{center}
\end{figure}

\subsection{X-ray properties \label{prop}}

Using the count rates obtained in the three energy bands mentioned in Sect.\,\ref{list}, we obtained hardness ratios (see columns 6 and 7 of Table\,\ref{xsrc}) defined respectively as
$$HR_1 = \frac{M - S}{M + S}$$
$$HR_2 = \frac{H - M}{H + M}.$$
We find that some sources present rather extreme hardness ratios, suggesting very hard X-ray spectra. Sources \#7, \#12, \#13, \#15, \#28, \#29 and \#31 have $HR_1$ values larger than 0.90 as shown in column 6 of Table\,\ref{xsrc} for EPIC-pn. We note that high hardness ratios are also obtained from EPIC-MOS instruments for these sources. Some of these sources may be extragalctic sources as discussed above.

We selected the brightest sources, and performed a spectral analysis of our EPIC data from the April and September observations.

Source \#1 is likely associated with the O6V((f)) star HD\,167834 (BD\,-12$^\circ$\,4969). This X-ray source was detected by the {\it EINSTEIN} satellite with a count rate of 0.012\,$\pm$\,0.004\,cts\,s$^{-1}$, leading to a flux\footnote{The flux was estimated assuming a thermal model with kT\,=\,0.5\,keV and a neutral absorption column obtained by integrating a model of the hydrogen particle spatial distribution in the Galaxy. The procedure is described in Grillo et al.\,(\cite{gril}).} of about 3.4\,$\times$\,10$^{-13}$\,erg\,cm$^{-2}$\,s$^{-1}$ (Grillo et al.\,\cite{gril}). If we take into account the exposure map correction, this source is the second brightest X-ray emitter in NGC\,6604, after HD\,167971 and before HD\,168112.

Source \#8 has a single counterpart in the three catalogues used for the source identification, and is most probably associated to the O9.5I or III star BD\,-12$^\circ$\,4982. No optical or infrared counterpart was found for source \#25. No information on the X-ray emission of these two objects has been found in the literature.

\begin{figure*}[ht]
\begin{center}
\resizebox{17cm}{6.0cm}{\includegraphics{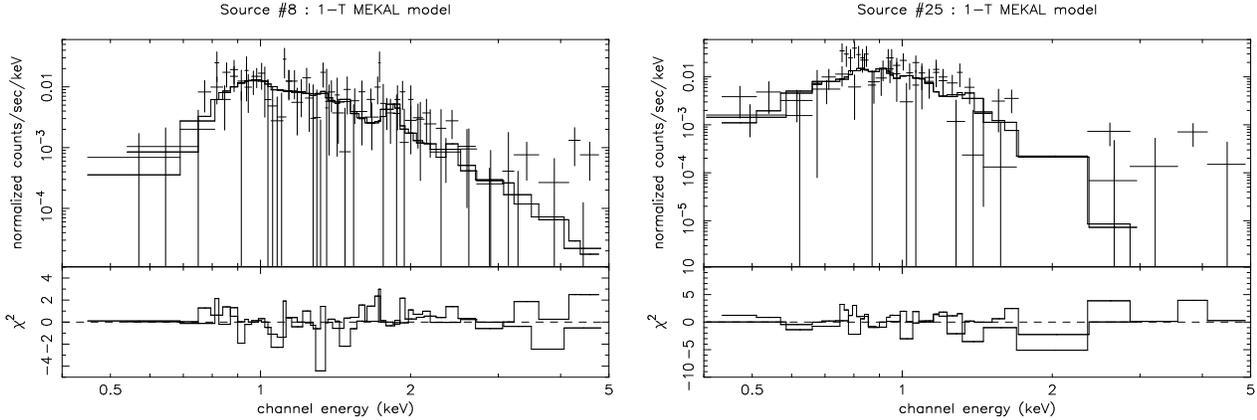}}
\caption{Combined EPIC-MOS1 and EPIC-MOS2 spectra of the September 2002 observation, fitted with a {\tt wabs$_\mathrm{ISM}$*wind*mekal} model for source \#8 ({\it left}) and a {\tt wabs*mekal} model for source \#25 ({\it right}). In the case of source \#8, the ISM absorption column was frozen at 0.55\,$\times$\,10$^{22}$\,cm$^{-2}$. \label{sp825}}
\end{center}
\end{figure*}

In the case of \#1, a 50\,$\arcsec$ radius circular region was used to extract the spectrum. The EPIC-pn spectrum was binned to have a minimum of 9 counts per channel. For \#8 and \#25, the spectra were extracted within a circular region with a radius of 40\,$\arcsec$. For the three sources, the background was extracted within an annulus centered on the source and covering the same area as the source region. Gaps and bad columns were excluded by rejecting properly adjusted rectangular boxes. The spectra were fitted following the same approach as described in Sect.\,\ref{res}. For sources \#8 and \#25 we obtained very similar fitting results from the EPIC-MOS and EPIC-pn spectra. All spectra from sources \#8 and \#25 were binned to reach a minimum of 5 counts per channel. The main results from the spectral analysis of these sources are summarized herebelow and in Table\,\ref{fitsrc}:
\begin{enumerate}
\item[-] \#1. From the ($B - V$)\,=\,0.95 color given by Forbes\,(\cite{for}) for this star, and the intrinsic color typical for an O6.5 star (Mihalas \& Binney\,\cite{MB}), we followed the same procedure as described in Sect.\,\ref{abs} to derive $E(B - V)$\,=\,1.26, yielding an ISM $N_\mathrm{H}$ of 0.73\,$\times$\,10$^{22}$\,cm$^{-2}$. We used the same ionized wind absorption model as for HD\,167971 to account for local absorption, and we left it as a free parameter. The EPIC-pn spectrum is best fitted with a two-temperature thermal {\tt mekal} model (see Fig.\,\ref{sp1}), with characteristic temperatures of about 0.2 and 1.4\,keV. The rather high temperature for the hotter component suggests that the intrinsic X-ray emission from the shocks in the stellar wind of the OB star (see e.g. Feldmeier et al.\,\cite{Feld}) is not the only process responsible for the observed X-rays. From this model, we derive an observed flux of $\sim$\,8.0\,$\times$\,10$^{-13}$\,erg\,cm$^{-2}$\,s$^{-1}$ between 0.4 and 10.0\,keV. Assuming that the distance to the star is 1.7\,kpc (Barbon et al.\,\cite{Bar}), we obtain an $L_\mathrm{X}$ corrected for the ISM absorption of 4.4\,$\times$\,10$^{33}$\,erg\,s$^{-1}$. If we consider the color excess derived above in this paragraph, the V magnitude from Table\,\ref{xsrc}, the distance to NGC\,6604 provided by Barbon et al.\,(\cite{Bar}), and the typical bolometric correction for an O6V star (Vacca et al.\,\cite{VGS}), we obtain a bolometric luminosity of about 1.7\,$\times$\,10$^{39}$\,erg\,s$^{-1}$. This yields an $L_\mathrm{X}$/$L_\mathrm{bol}$ ratio of 2.6\,$\times$\,10$^{-6}$. Considering the X-ray luminosity expected from the empirical relation of Bergh\"ofer et al. (\cite{BSDC}), e.g. $\sim$\,2.8\,$\times$\,10$^{32}$\,erg\,s$^{-1}$, we obtain an X-ray luminosity excess of about 16! We mention that if we use the typical bolometric luminosity for a single O6V star (i.e. 9.8\,$\times$\,10$^{38}$\,erg\,s$^{-1}$, Howarth \& Prinja\,\cite{HP}), HD\,167834 appears a factor 30 overluminous in X-rays.  The spectrum of this source is also compatible with a model including a thermal component and a power law, but the quality of the fit is somewhat poorer than that obtained for the purely thermal model (see Table\,\ref{fitsrc}). The possible explanations to produce additional X-rays associated to a plasma temperature of a few 10$^{7}$\,K could be an interaction between the winds of two OB stars in a binary system (see e.g. Stevens et al.\,\cite{SBP}), or possibly a magnetic confinement scenario as proposed by Babel \& Montmerle\,(\cite{BM}). However, the huge luminosity excess can more probably be explained by the binary scenario. Higher quality data are needed to discuss the physical origin of this X-ray emission.
\item[-] \#8. We adopted the same approach as for source \#1 to establish the value of the ISM hydrogen column density. With a ($B - V$)\,=\,0.64 (Forbes\,\cite{for}), we derive an ISM $N_\mathrm{H}$ of 0.55\,$\times$\,10$^{22}$\,cm$^{-2}$. A second absorption column was left as a free parameter to account for the absorption by the wind material. The best fit is obtained between 0.4 and 5.0\,keV with a single temperature {\tt mekal} model, with a characteristic temperature of about 0.7\,keV. We do not find any significant difference in the model parameters obtained for our two observations. We obtain an oberved flux between 0.4 and 5.0\,keV of 7-8\,$\times$\,10$^{-14}$\,erg\,cm$^{-2}$\,s$^{-1}$. The fact that this source presents a rather soft spectrum is consistent with the rather modest hardness ratios reported in Table\,\ref{xsrc}. The combined fit of EPIC-MOS spectra is shown in Fig\,\ref{sp825}.
\item[-] \#25. As no optical counterpart was found for this source, we were not able to determine a priori an ISM hydrogen column density in its direction. So we used a unique absorption column which was left as a free parameter. The spectral analysis of this source between 0.4 and 5.0\,keV reveals a very soft spectrum fitted with a one-temperature model with a kT of about 0.2 keV, which is very consistent with the $HR_1$ and $HR_2$ values quoted in Table\,\ref{xsrc}. We did not find any indication for a significant variability of the model parameters between our April and September observations. Figure \ref{sp825} shows the EPIC-MOS spectra fitted simultaneously with the 1-T model. 
\end{enumerate} 
 
\subsection{X-ray variability}
We ran the {\tt edetect\_chain} metatask on the three EPIC data sets simultaneously, separately for the two observations, and we carefully inspected the images in both cases to check whether the sources listed in Table\,\ref{xsrc} are detected at both epochs. Most of the apparent variability observed for the X-ray sources can be explained by the fact that the effective exposure times of our two pointings are significantly different ($\sim$\,6\,ks in April and $\sim$\,11\,ks in September for EPIC-pn, and $\sim$\,9\,ks in April and $\sim$\,13\,ks in September for EPIC-MOS), and to some extent can also be due to gaps or bad columns. However, it appears that the sources \#19 and \#21 are only detected in September 2002, whilst sources \#15, \#28 and \#31 are detected only in April 2002. This strong variability suggests a flaring behaviour. We extracted light curves to investigate their variability with more details, but these sources are too faint to provide conclusive results. Source \#31 presents apparently a sufficiently high count rate to give a valuable light curve, but we note that this source lies very close to the border of the EPIC field. Consequently, the observed count rate is about a factor of 3 lower than the exposure corrected value quoted in Table\,\ref{xsrc}. Among these targets, only \#15, \#28 and \#31 present a very high hardness ratio $HR_1$. A comparison of the count rates obtained from the two separate observations reveals no significant variability for the other X-ray sources detected with {\it XMM-Newton}.

We finally searched for a short time scale variability in the cases of the X-ray sources discussed in Section\,\ref{prop}, i.e. \#1, \#8, and \#25. We extracted light curves with time bins ranging from 100\,s to 1000\,s following the same approach as used in the case of HD\,167971 (see Sect.\,\ref{var}). Variability tests (chi-square, {\it pov}-test)  performed on these light curves did not reveal any significant variability during our two {\it XMM-Newton} observations for any of the three EPIC instruments.

\subsection{Star formation in NGC\,6604? \label{form}}

As mentioned in the previous subsections, several indicators (possible flaring activity, position in the $JHK_S$ diagram...) suggest some of the X-ray emitters in NGC\,6604 may be candidate PMS stars (Feigelson \& Montmerle\,\cite{FM}). This fact is at first sight contradictory to the statement of Barbon et al.\,(\cite{Bar}) that pre-ZAMS objects are not detected in NGC\,6604. However, this lack of detection could be partly explained by the fact that the study of these authors was limited to an area of only 2.1\,$\times$\,3.3 arcmin$^{2}$ centered on HD\,167971. Most of the X-ray sources we detect are indeed located outside this area. 

The fact that our investigation is only able to reveal the brightest X-ray emitters harboured by NGC\,6604 could strongly bias our investigation of PMS activity in the cluster. Let us consider the faintest X-ray sources of our list. Assuming a simple thermal model with a kT of about 0.5 keV\footnote{With an absorbing interstellar column of about 0.6\,$\times$\,10$^{22}$\,cm$^{-2}$, i.e. an approximate mean value of those used for the spectral analysis of HD\,168112 and HD\,167971.}, we obtain an unabsorbed flux between 0.4 and 10.0\,keV of about 1.0\,$\times$\,10$^{-14}$\,erg\,cm$^{-2}$\,s$^{-1}$. For a distance to the cluster of about 1.7\,kpc (Barbon et al.\,\cite{Bar}), this unabsorbed flux leads to a luminosity for the faintest source detected in our data of about 3.5\,$\times$\,10$^{30}$\,erg\,s$^{-1}$. This value is of the order of the highest typical luminosities expected for PMS stars when no flare is occuring (Feigelson \& Montmerle\,\cite{FM}). This suggests that our data only allow a detection of the more luminous PMS candidates within NGC\,6604, and possibly during flaring stages, i.e. at their highest X-ray emission state. Deeper X-ray observations are needed to investigate the PMS populations in NGC\,6604.

Unlike the situation in NGC\,6383 (Rauw et al.\,\cite{ngc6383}) and NGC\,6231 (Sana et al.\,\cite{ngc6231}), the X-ray sources in NGC\,6604 reveal no concentration around the more massive cluster members. In NGC\,6383 and NGC\,6231, the fainter X-ray sources are mainly associated to PMS stars, thereby suggesting a close relationship between the formation of low-mass and high-mass stars. In NGC\,6604, the rather sparse spatial distribution of the members of the cluster already noted in the visible domain, show that this is not a mass-segregated open cluster. Following the $N$-body simulations performed by Bonnell \& Bate (\cite{BB}), gas accretion should indeed have led to a strong contraction of the cluster responsible for a significant mass segregation. A possible explanation for the absence of notable mass segregation in open clusters was proposed by Vine \& Bonnell (\cite{VB}). According to these authors, the strong stellar winds of O and B stars in young open clusters result in a significant gas removal from the core, where the OB stars are preferentially formed. Their simulations show that the changes in the gravitational potential of the core as a result of this gas expulsion is able to break the mass segregation  ({\it core dissolution}), allowing stars to drift away from the core. The apparently different dynamical stages between NGC\,6604 and NGC\,6383 is not unexpected regarding their ages, i.e. respectively 5\,$\pm$\,2 (Barbon et al.\,\cite{Bar}) and 1.7\,$\pm$\,0.4\,Myr (FitzGerald et al.\,\cite{fitz}).

\section{Summary and conclusions \label{sect_concl}}

The analysis of our {\it XMM-Newton} data from HD\,167971 reveals that the soft part of the X-ray spectrum is thermal with typical plasma temperatures of about 0.2 to 0.8 keV, i.e. 2\,$\times$\,10$^{6}$ to 9\,$\times$\,10$^{6}$ K. The harder part of the spectrum is fitted equally well with a thermal or a power law component. On the one hand, if the higher energy emission is thermal, the plasma temperatures ($\sim$ 2.3\,$\times$\,10$^{7}$ to 4.6\,$\times$\,10$^{7}$ K) are compatible with those expected for a shock heated plasma in a wind-wind collision zone, where the winds have reached speeds close to their terminal velocities. If we consider the short period of the close binary system, such velocities are not expected at the position of the collision zone of the winds of the O5-8V components. Provided the three components are indeed physically connected, this argues in favour of an interaction with the third more distant companion. On the other hand, if the hard energy emission component is non-thermal, the photon index is rather high ($\sim$ 3), which is reminiscent of the case of other non-thermal radio emitters like 9\,Sgr (Rauw et al.\,\cite{9sgr}) or HD\,168112 (De Becker et al.\,\cite{paper1}). The X-ray luminosity exceeds the value expected for the `canonical' $L_\mathrm{X}/L_\mathrm{bol}$ relation by a factor 4. This excess is larger than that found for HD\,168112 (De Becker et al.\,\cite{paper1}). Finally, we report on a weak decrease of the X-ray flux between our two observations separated by about five months possibly due to an eclipse of the O5-8V + O5-8V close binary system.\\

By considering several results from recent observations of massive stars with X-ray satellites ({\it XMM-Newton}) and radio telescopes (e.g. VLA), we note that the non-thermal radio emitters are possibly not the best candidates to display a non-thermal emission in the bandpass of current X-ray observatories like {\it XMM-Newton} or {\it Chandra}. In a scenario where non-thermal radio emission requires a wind-wind collision in a binary system, wide binaries produce too hard a thermal spectrum to allow the detection of any putative hard non-thermal contribution. On the contrary, short period binaries are not expected to produce an observable synchrotron radio spectrum because of the huge opacity of the wind material for radio photons. They are however likely to produce a softer thermal X-ray spectrum than long period binaries, offering the possibility to unveil the expected hard non-thermal X-ray emission component. In other words, the simultaneous detection of a non-thermal emission in the radio and X-ray domains seems rather unlikely. More X-ray and radio observations of massive binaries (O and WR), whatever the thermal or non-thermal nature of their radio emission, are needed to interpret the non-thermal emission from massive stars.\\

Finally, our investigation of the X-ray emission from NGC\,6604 led to the detection of 29 sources in addition to HD\,168112 and HD\,167971. The cross-correlation of the position of these sources with the GSC, the 2MASS and the USNO catalogues revealed that most of them have at least one optical or infrared counterpart. We describe a more complete spectral analysis in the case of the three brightest  objects of our source list. Two of them are associated to BD\,-12$^\circ$\,4982 (O9.5I or III) and HD\,167834 (O6.5V((f))) respectively, and the third one has no known optical or infrared counterpart. Some indicators of stellar formation activity suggest a few sources of our list may be PMS objects. Unfortunately most of the sources detected in NGC\,6604 are too faint to be investigated in details in the context of this study. Finally, the X-ray emitters of NGC\,6604 detected with {\it XMM-Newton} are not concentrated close to the most massive stars of the cluster. This rather sparse spatial distribution contrasts with that observed in other open clusters like NGC\,6383 and NGC\,6231 where young stellar objects detected in X-rays are mainly concentrated close to the most massive stars. 

\acknowledgement{Our thanks go to Alain Detal (Li\`ege) for his help in installing the {\sc sas}, to Hugues Sana for helpful discussions on the X-ray source detection procedure, and to Ya\"el Naz\'e for providing the wind absorption model routines and for helpful discussions. The Li\`ege team acknowledges support from the Fonds National de la Recherche Scientifique (Belgium) and through the PRODEX XMM-OM and Integral Projects. This research is also supported in part by contract P5/36 ``P\^ole d'Attraction Interuniversitaire'' (Belspo). This research has made use of the SIMBAD database, operated at CDS, Strasbourg, France and NASA's ADS Abstract Service. This publication makes use of data products from the Two Micron All Sky Survey, which is a joint project of the University of Massachusetts and the IPAC/California Institute of Technology, funded by NASA and the NSF.}

\end{document}